Kurdistan Regional Government
Ministry of Higher Education and Scientific Research
University of Sulaimani
College of Agricultural Engineering Sciences

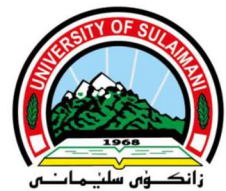

# EFFECT OF SEAWEED, MORINGA LEAF EXTRACT AND BIOFERTILIZER ON GROWTH, YIELD AND FRUIT QUALITY OF CUCUMBER (*Cucumis sativus* L.) UNDER GREENHOUSE CONDITION

**A Thesis**

**Submitted to the Council of the College of Agricultural Engineering Sciences at the University of Sulaimani in Partial Fulfillment of the Requirements for the Degree of Master of Science**

**in**

**Horticulture**

**(Organic Farming)**

By

**Roshna Faeq Kakbra**

B.Sc. in Horticulture College of Agricultural Engineering Sciences, University of Sulaimani (2012)

Supervisors

**Dr. Ghuncha Kamal Tofiq**          **Dr. Salam Mahmud Sulaiman**
Lecturer                             Assistant Professor

2723 K                                                            2023 A.D

# Summary

This factorial experiment was conducted in a greenhouse during the period of May 3, 2021 to August 5, 2021 at the research farm belongs to the Horticulture Department, College of Agricultural Engineering Sciences, University of Sulaimani, Sulaimani, Iraq. The experiment was designed to study the effect of some biostimulants, individually and their combinations, on cucumber plants performance under greenhouse conditions; in addition to compare the results with 100% of chemical fertilizers application. The treatments consisted of control (without adding any kinds of biostimulants) recommended dose of 100% chemical fertilizers (RDCF), seaweed extracts (SE), moringa leaf extract (MLE), bacterial-based biostimulant of Fulzym- plus (FP), that contains *Bacillus subtilis* and *Pseudomonas putida*, (SE+MLE), (SE+FP), (MLE+FP), and (SE+MLE+FP). The experiment was laid out in simple – RCBD with 3 replications.

The results showed that the application of different biostimulants, individually and their combinations, significantly improved the root growth characteristics at $P \leq 0.05$. However, the highest values of lateral roots number per plant, lateral root length, lateral root diameter and root system dry weight were recorded by the application of recommended dose of chemical fertilizer (RDCF). While, this treatment was not different substantially with the triple combination of the tested biostimulants (SE+FP+MLE) in all studied root characteristics. In addition, untreated plants registered the minimum value of all the mentioned characters.

In addition, majority of the biostimulant treatments showed positive effective on the vegetative growth characteristics compared to untreated plants, but the triple combination between them (SE+MLE+FP) was more affective. Even though the chemical fertilization treatment (RDCF) recorded the longest plant shoot, highest number of nodes per plant, thickest stem, maximum plant leaf area, highest percentage of leaf dry matter and highest intensity of chlorophyll pigment in the leaves. At the same time, this treatment was not various markedly with the co-application of the biostimulants (SE+MLE+FP) in all mentioned traits related to the vegetative growth. Also, control plants recorded minimum values for all vegetative measurements.

Furthermore, applying the biostimulants, individually and their combinations, had a substantial impact on improving the concentration of macronutrients (N, P and K) and micronutrients (Fe and Zn) if compared to control, with exception to the individual effect of (MLE) and (FP) on Fe concentration as well as dual application between (SE+FP) on Zn concentration. Despite the recommended dose of chemical fertilizer (RDCF) resulted in the highest concentration of N, P and Fe; and applying bacterial-based biostimulant of FP obtained the highest concentration of K and Zn. While, these treatments were not different statistically with the treatment of (SE+FP+MLE) in all measured nutrients. The control treatment recorded the minimum concentration of all the measured nutrients.

# CHAPTER ONE

# INTRODUCTION

Cucumber (*Cucumis sativus* L.), belongs to the Cucurbitaceae family, and is one of the most popular vegetable crops throughout the world. It is divided according to their uses, either for pickling or for fresh consumption purposes (Kaur and Sharma, 2021). Cucumber fruits have a distinctive flavor that increased their economic value as a vegetable crop. This crop is planted annually during the spring and fall seasons in the open fields and greenhouses conditions (Diab et al., 2016). In Kurdistan Region, the number of greenhouses reached about 29861 houses in the end of 2021, which most of them located in Sulaimani governorate. Approximately, 90% of these houses are cultivated annually with the cucumber crop (Ministry of Agriculture and Water Resources, 2021).

Usually, the traditional or conventional method is used for cucumber production inside these houses, in which many farmers are using high amounts of chemical fertilizers in order to get higher productivity. It is known that the excessive utilization of chemical fertilizers causes negative impacts on the environment and human health, in addition to waste a tremendous amount of money. Therefore, one of the suggested alternative approaches to produce healthy fruits without adverse of the environment is organic agriculture. Furthermore, increasing consumer awareness concerning healthy food favors the enhancement of the importance of organic farming (Drobek et al., 2019). In the last decade, several modern methods were used to enhance production of the vegetable crops organically through the usage of natural or organic substances and beneficial microorganisms as plant biostimulants. According to the European Biostimulant Industry Council; plant biostimulants are biological origin, with microorganisms, can applied to soil (roots) or plants which is stimulate physiological processes, promoting and increased growth, production quality with greater tolerance to stress (EBIC, 2013).

Seaweed extracts (SE), particularly the brown macro-algae, are often used as biostimulants because of their macro content and micro nutrients, phytohormones actives and they have also role in signaling molecules such as polysaccharides, and betaines, which enhance plant growth, yield and fruit quality without adverse environmental impacts (La Bella *et al.,* 2021). Increasing plant production under unfavorable circumstances is achieved by altering a number of biochemical and physiological mechanisms, including the stimulation of enzyme activities involved in glycolysis, phytohormones and Krebs cycle, additionally it boosts of minerals approval and accumulation of treated plants through modifications morphological of root (Colla *et al.,* 2017; La Bella *et al.,* 2021).

The plant *Moringa oleifera*, commonly known as drumstick tree or horseradish, was characterized as a source of secondary metabolites different compounds with biostimulant potential. In particular, moringa leaf extracts (MLE) contains macro and micronutrients, amino acids, and vitamins. Additionally, it is rich in secondary metabolites compounds such as; phytohormones and antioxidants (Gopalakrishnan *et al.,* 2016; Brazales-Cevallos *et al.,* 2022). The MLE has been widely used in organic agriculture to improve plant growth and development as well as enhancing fruit quality characteristics of various vegetable crops.

Bacterial-based biostimulants contain a main group of plant biostimulants. Plant growth-promoting rhizobacteria (PGPR) that colonize the plant rhizosphere are the most prominent group in this category. The illustrative advantageous groups of PGPR include *Pseudomonas* spp., and *Bacillus* spp. These bacteria improve plant growth, enhance nutrients uptake in plants, and tolerance to biotic and abiotic stresses (Hamid *et al.,* 2021; Radhakrishnan *et al,* 2017).

The knowledge mentioned above contributes to the promotion of using the biostimulants in agricultural practices as environmentally-friendly alternatives to traditional crop management strategies, which have mostly depended on the application of synthetic agrochemicals. In Kurdistan Region, low knowledge and limited published data is available regarding the effect of different plant biostimulants on cucumber plant performance under protected conditions. Therefore, the aims of this study are to investigate the impact of SE and MLE as non-microbial biostimulants as well as bacterial-based biostimulant of Fulzym-plus (FP) that contains *Bacillus subtilis* and *Pseudomonas putida*, separately and to their combinations on the nutrient's status, reproductive growth, yield and fruit quality characteristics of cucumber plants grown under greenhouse conditions.

# CHAPTER TWO

# LITERATURE REVIEW

## 2.1 Cucumber and its Nutritional Importance

One of the greatest popular vegetable crops which belongs to the gourd family (Cucurbitaceae) is Cucumber (*Cucumis sativus* L.). It is an important crop in terms of economical and nutrition value worldwide, which is grown in open fields or under greenhouse conditions (Pal *et al.*, 2020). It is regarded as a vital vegetable for fresh consumption as a food or medicinal since ancient time in India and the cucumber was domesticated ~3000 years ago (Paris *et al.*, 2012). Abbey *et al* (2017) stated that there are several varieties of cucumber but the edible ones can be classified into two groups: the slicing cucumbers and pickling cucumbers.

Cucumber is a native to south Asia, specifically in the warm and humid climate of the Himalaya in north west India and probably northern Africa (Pal *et al.*, 2020). According to the last available statistics in 2019, cucumber was ranked $10^{th}$ among the most important vegetable crops worldwide. It was globally cultivated on an area of 2,231,402 hectares and yielded around 87,805,086 tons. In addition, the top 10 producers of fresh cucumber in the world in 2019 are China, Turkey, Russia, Ukraine, Iran, Uzbekistan, Mexico, Spain, United States and Japan. China is the global leader in cucumber production that shared 70,338,971 tons (80.11%) in global production from 1,258,370 ha (56.39%) of cultivation area (FAOSTAT, 2021).

Regarding the nutritional value, each 100 g of fresh weight of cucumber contains 12-15 cal energy, 0.6-0.65 g protein, 0.1-0.11 g total lipids, 2.2 -3.63 g carbohydrates, 0.28-0.3mg iron, 14-16 mg calcium, 2.8-12 mg vitamin C, and 1.125-5 μg vitamin A (β-carotene). In addition, other vitamins such as vitamin K, B6, B12, thiamine, niacin, as well as phenols, flavonoids, and triterpenes, which are able to act as antioxidant and anti-inflammatory agents (Aderinola *et al.*, 2019). More studies confirmed that cucumber has also several health benefits such as: re-hydrating the body, regulating blood pressure, cholesterol reduction, body weight management, cancer prevention, bone health, diabetes cure and antioxidant activity (Kumar *et al.*, 2010; Bello *et al.*, 2014 and Uthpala *et al.*, 2020).

Furthermore, there are several factors that have a direct impact on the nutritional value of cucumbers such as varieties and agricultural practices including fertilization. Many literatures indicate that organically produced cucumber have much higher nutritional values than chemically grown cucumbers (Ramakrishnan *et al.*, 2021 and Adekiya *et al.*, 2022).

## 2.2 Overview of Organic Farming

worldwide people will grow from 7.4 to 9 billion by 2050, which is expected to increase the demand for food by 70%. Therefore, people are worried about the sustainability of agriculture and health but fortunately has significantly evidence for that about recent study from Research Institute of Organic Agriculture (FiBL) on organic farming land international that data was collected unite the end of 2018 from 186 countries (Willer *et al.,* 2020).

Production of crop systematically and scientifically acknowledged without utilized of non-natural growth regulators, chemical fertilizers, pesticides and livestock food flavors known as organic farming. Historically, first time famous scientist Northbourne at his name book ('Look to the Land') at 1940 was describe the term of organic agricultural, which is "the farm unit must be have a living entity; biological completeness and balanced organic life." Northbourne 2003 also described it as "management of ecosystem that encourages with progress activity of biological cycles and soil biodiversity". In conventional agricultural, chemicals and artificial pesticides are used to remove weeds, pests and insects, so as to increase developed plant production and income artificial hormones and fertilizers are used as well (Worthington, 2001). Therefore, consumers are always concern about conventionally grown foods, also increasing consumer demand for organic food have significant goals, so as to enhance the farmer to produce the organic food, consequently also may have significant influence on both social health and environment (Das *et al.,* 2020).

Generally, organic agriculture is a collection of agricultural practices system for the agronomy of crops and childhood of animals in normal habits which can be regenerates the public health, improved ecosystem. organic farming was described by Food and agriculture organization of the United Nation (FAO) as a "Organic farming is a unique production management system which can enhances the welfare of agro-ecosystem health including: biodiversity, soil biological activity, using agronomic, biological and mechanical approaches can applied on farm where synthetic off-farm inputs are excluded" (Meena *et al.,* 2013).

According to United States Department of Agriculture (USDA) organic agricultural is described as a "production system where synthetic compounds like pesticides, fertilizers and hormones are totally avoided or excluded, while other synthetic free measures such as, crop rotations, crop residues and animal manures are applied for plant nutrition and protection" (Meena *et al.,* 2013).

In 2019, 72.3 million hectares were super-visioned below organic farming around the world. The regions with the main areas of organic farming field are Oceania (36 million hectares, which is half the world's organic agricultural land) and Europe (16.5 million hectares, 23 percent). Latin America has 8.3 million hectares (11 percent) followed by Asia (5.9 million hectares, 8 percent), North America (3.6 million hectares, 5 percent), and Africa (2 million hectares, 3 percent). In addition, organic agriculture was practiced in 187 countries, and 72.3 million hectares of agricultural land were managed organically by at least 3.1 million farmers. The global sales of organic food and drink reached more than 106 billion euros in 2019 (FAOSTAT, 2021).

## 2.3 Problems of Inorganic or Chemical Fertilizers

In the last decade, growing and demands for food may play an important factor to increase yield per unit area in crop production. While, intensive use of chemical fertilizer during and post the green revolution caused a number of environmental and ecological problems such as soil acidification and degradation, water eutrophication, global warming and damage of ecosystem (Savci, 2012 and Lu and Tian 2017). Regarding the human health, many studies demonstrated that accumulation of toxic substances in the agricultural products, including nitrates, sulfates, and heavy metals; after utilized of chemical fertilizers for long-term which is finally causes threatens human health during food consumption (Sun *et al.,* 2017 and Li *et al.,* 2021).

## 2.4 Plant Biostimulants and Their Role in Organic Farming

One of the vital and more sustainable methods in modern agriculture is plant biostimulants (PBs) which is a portion of an integrated crop management (ICM) system (Hamid *et al.,* 2021). In the last decade, PBs have been defined by authors based on the source material, structure of the products and mode of action. PBs is any biological materials or microorganisms naturally and can be designate commercial products that can be increased into plants to facilitate the nutrient uptake, increase crop quality and tolerance to abiotic stress (Du Jardin, 2015). In addition, fertilizing products regulation (FPR) defined PBs as "Products stimulating plant nutrition processes independently of the product's nutrient content with the sole aim of improving plant rhizosphere: nutrient use efficiency, resistance agents to abiotic stress, quality traits, availability of limited nutrients rhizosphere zone" (European Parliament and European Council, 2019). Generally, bio-stimulants categorize was described by many scientistic that the last classified substances depending on (Torre *et al.,*2016) described categories of bio-stimulants consist of types including; Humic materials, seaweed, Hydrolyzed proteins and amino acids, Inorganic salts and Microorganisms.

Regarding above-described, plant biostimulants are amongst the hot subjects in agriculture which have been widely studied because of its promising environmental-friendly (Sangiorgio *et al., 2020*; Baltazar *et al.,* 2021 and Castiglione *et al.,* 2021). In the twenty-first century, role of plant biostimulants in organic farming have been demonstrated that they are new novel alternative method in modern farming that eco-friendly of ecosystems and a major facing to get sustainability of foods (Shubha *et al.,* 2017). They have many benefits including, improve nutrient uptake, increase the surface contact between roots and soil, bioavailability of soil nutrients by secreting enzymes (i.e., phosphatases) (Calvo *et al.,* 2014 and Colla *et al.,* 2015a). In addition, they can be extensively utilized agricultural practices, which has implications of numerous benefits to encourage plant development and resistant against stresses (Hassan *et al.,* 2017). Furthermore, performs of organic farming seek advanced solutions concerning on quality and quantity of plant production, by increasing nitrogen use efficiency (NUE), availability and adjusted of nutrient (Barbieri *et al.,* 2015; Reganold, 2016 and De Pascale *et al.,* 2017).

Albrecht, (2019) introduced the use of artificial biostimulants marked was estimated about 2.19 billion in 2018 around the worldwide, and also may significantly predictable the twelve-monthly development rate of 12.5% from 2019 to 2024. In EU can nearly 40% of the market share which is the largest marketplace for reproduction of biostimulants, while in the North American shops is assessed to rate $605.1 million in 2019.

Furthermore, the global biostimulants (BSs) market continues to grow rapidly, surpassing €2.7 billion in 2022, propelled by many governments' increasing focus on improving sustainability while reducing the environmental footprint of food production (Critchley *et al.,* 2021). In recent years, advancement in biochemical, genomic, and transcriptomic tools significantly contributed to unveiling the mode of actions of BSs (Bulgari *et al.,* 2017 and Abbott *et al.,* 2018). This advancement has opened the doors for many BSs related industries to look for more effective and reliable formulations by blending microbial BSs with non-microbial BSs.

### 2.4.1 Non-microbial plant biostimulants

Non-microbial biostimulant is one of the most categories of plant biostimulants that can be achieved naturally or artificially including: seaweed extracts, chitosan, humic substances, protein hydrolysates, nitrogen-containing compounds and inorganic compounds. Humic and fulvic acids are developed from the biological or chemical conversion of plant and animal wastes (Jindo *et al.,* 2020). Also, in agronomy sector protein hydrolysates as PBs are usually utilized to assistance plants manage with tolerance to stresses, because of oligo and polypeptides with a mixture of free amino acids, which can be derived from PBs (Rouphael and Colla, 2018). The seaweeds are highly in polysaccharides, enzymes, proteins and highly nutritious. Therefore, in section below will specifically explain seaweed extracts.

### 2.4.1.1 Seaweed extracts (SEs)

Seaweed extracts (SEs) represent an important category of organic non-microbial PBs, which is defined as a biostimulants and consist nearly10,000 species of macroalgae. it also is estimated 10% of marine production around the world. Depending on pigments red, green, and brown macroalgae are the most common SWE that utilized in farming with numerous commercial products present in the market (Khan, *et al.,* 2009 and Battacharyya *et al.,* 2015).

Seas and oceans are a renewable source to get macro algae as PBs (Rouphael and Colla, 2020). Most populations of algae are originated in aquatic habitats however, algae can occupy both terrestrial (soil and stone) habitats and aquatic (fresh and marine waters) (Kurepin *et al.,* 2014). Alga 600 is a natural organic seaweed, and commercially used as 'natural' biostimulators of plant growth and development that unique extract from brown seaweed such us; Sargassum, *Ascophyllum Nodosum* and *Laminaria*, Brown algal genus *Sargassum* (viz. Sargasso Sea) is a great biomass species among three main species (brown, red marine and green macroalgae) that could be free-floating (Battacharyya *et al.,* 2015). In addition, brown seaweeds extracts are extensively utilized in production of horticulture as a plant growth-promoting and to abiotic stresses resistance such as; dangerous temperatures, drought, nutrient deficiency and salinity. Extract of seaweed contain many chemical compounds such as fatty acids, mineral nutrients, complex polysaccharide, vitamins, and phytohormones (Ali *et al.,* 2021).

Seaweed extracts are usually utilized in sustainable agriculture. Therefore, in this present study are mostly focused on bioactive and elicitor components of seaweed extracts with mechanisms and mode of seaweeds as a bio-stimulants activities, and effect in production of vegetables particularly fresh cucumber in term quality and quantity. Generally, a major concern to the advancement of agriculture is biotic and abiotic stress, which has directly impact on the plant health and production. Therefore, spray seaweed extracts on plants can enhanced resistance to salt and freezing abiotic stresses (Mancuso *et al.,* 2006). However, improving and priming effects on plant's tolerance against both abiotic and biotic stresses, via seaweed extracts can be credited depend on chemical structure of the extracts and eliciting properties as well as (Yakhin *et al.,* 2017). In addition, treated seaweed extracts with plants including, A. nodosum and Sargassum spp. may able to survive the damaging impact of these abiotic stresses. Researcher was also reported that tomato plant can treated with seaweed extracts, led to reduction leaf osmotic potential and preventing extensive damage (Wilson, 2001). Furthermore, Seaweed extracts has been significant role during severe drought, cold, and salinity stress on plant by enhanced root phenotype, a build-up of non-structural carbohydrates that enhanced storage of energy, improved metabolism, and water adjustments, build-up of proline (Dalal *et al.,* 2019 and Ganesan *et al.,*2015).

### 2.4.1.2 Effect of SEs on improving plant growth, yield and fruit quality of vegetable crops

Seaweed extracts (SEs) is type of biostimulant extracted that can enhance plant growth development, progress crop quality, increase tolerance to stress, and beneficial to soil improvement (Tudu Keya *et al.,* 2022), because of it is rich in macro and micro elements, with vital plant hormones such us; cytokinin, gibberellins and auxins. Researcher demonstrated that those hormones lead to enlargement, increasing and division of cell consequently, improving growth parameters with increasing photosynthesis processes (Layek *et al.,* 2018).

In addition, seaweed, substances responding a depend on many factors such as: application time, extract concentration, and application methods, like foliar, root, fertigation, drenching, and adding to the soil or growing medium (Jayaraman and Ali, 2015). Furthermore, seaweed extracts have been extremely used in agricultural production for improving biomass yield in term quality and quantity that can be have significant effect on seed germination and plant development during harvest and post-harvest as well (Ali *et al.,* 2019). Many researchers have proven that the positive impact of this extract including; Rayorath *et al.,* (2008) demonstrated that seaweeds can help increased rates of germination and have important growths in seedling vigor by promoting root size and mass and also can defend seedlings from transplantation shock in cabbage, marigold and tomato (Crouch and Van Staden, 1993). Promote plant growth have been conducted by marine algae, for instance; seaweed-based extract application on greenhouse-grown bean (*Phaseolus vulgaris* L.) and silver beet (*Beta vulgaris* L.) plants were shown to increase yield and dry weight accumulation (Kurepin *et al.,* 2014). Trejo Valencia *et al.,* (2018) was concluded that positive effects on the plant growth, fruit yield, and quality of cucumber fruits was confirmed by foliar application with seaweed extract. Furthermore, Ahmed and Shalaby (2012) presented that the utilized of commercial seaweed extracts with E. intestinelis, G. pectinutum, can improve vegetative development and cucumber plant yield.

Many scientists also reported that significant increase of plant yield with nutrient quality were observed after using seaweed extracts in different economical crops such as: cucumber, tomato, pepper, lettuce, spinach, and strawberry (Xu and Leskover, 2015; Ali *et al.,* 2016; Trejo Valenci *et al.,* 2018; Kapur *et al.,* 2018 and Yusuf *et al.,* 2019). In addition, Hassan *et al.,* (2021) demonstrated that the foliar application of seaweed (TAM) can significantly increase the yield, growth and quality of cucumber (*Cucumis sativus*) as comparison with use of conventional NPK fertilization. Moreover, brown seaweeds extracts are widely used in manufacturing of vegetables a plant growth-promoting and to abiotic stresses resistance including; drought, dangerous temperatures, nutrient deficiency and salinity, physiological impacts of brown seaweed extracts were also reported on vegetable crops, and concluded that numerous of beneficial including; nutrient uptake improved, tolerance to biotic and abiotic stress, and over all vegetable production (Battacharyya *et al.,* 2015).

Plant biostimulants were used to developments of cucumber plants under soilless culture, consequently, many traits of cucumber crops including (shoot and root fresh weight, stem width, height, number of leaf and area) were positively improved (Abdelgalil *et al.,* 2021). Researcher was also suggested that apply of bio-stimulants safely and environmentally friendly alongside of a significant consequence on growing and development of cucumber production crops. Commercial extracts of a wide variety of seaweeds are increasingly popular, not only for use as biostimulants of plant growth, but also to impart tolerance to several abiotic, environmental stresses (Sangha *et al.,* 2014). In addition, significant effect due to seaweed extract Aly (sea force1) on stem length, plant dry weight, chlorophyll, carbohydrates, fruits diameter and length, early and total yield, fruit firmness, total soluble solids and vitamin C contents in fruits and nitrogen, phosphorous and potassium in leaves of cucumber plants (Abdulraheem, 2009).

Boukhari *et al* (2020) demonstrated that increased yield, stimulated the plant growth, uptake of soil nutrients by plants can observed after added SEs exract. On the other hand, soil SEs applications can improve of metabolic activities of soil microbes and the colony counts consequently, donated to rise plant shoot growth and root (Alam *et al.*, 2013). For instance, researchers confirmed apply SEs on onion crop grown under water stress can significantly improve N, P, and K minerals by 116, 113, and 93% respectively as compared to the control plants (Almaroai and Eissa, 2020). Another study found that apply SEs to *Brassica napus* L. can up-regulated genes associated with photosynthesis, cell metabolism, stress response and S and N metabolism, and also increased the biogenesis of chloroplasts due to increased chlorophyll content by reducing degradation of chlorophyll (Jannin *et al.,* 2013). During, vegetative stage chlorophyll content of leaves sweet pepper and tomato was increased by application of *Ascophyllum nodosum* which was also probably due to inhibition of chlorophyll degradation caused partly by betaines present in the extract (Blunden *et al.,* 1996). Inhibition of chlorophyll degradation may be referred to betaine compounds in the seaweed extracts which was let to photosynthetic activity (Genard, *et al.,* 1991).

## 2.4.1.3 Effect of SEs on nutrients and phytochemical contents of the vegetable plants

SE has been utilized in agriculture for many years, which is mainly comprise natural hormones, substance including; gibberellin, abscisic acid, auxin, cytokinin and active substances such as; Seaweed extracts contain a wide variety of plant growth-promoting substances such as gibberellins, cytokinins, betaines, auxins, and organic substances, including amino acids, macronutrients that increase crop yield and quality (Abd El Moniem, and Abd-Allah, 2008; Sathya *et al*., 2013). Phenolic and betaine compounds, sugar alcohol and seaweed polysaccharide (Delaunois *et al.,* 2014 and Battacharyya *et al.,* 2015). It can elicit and directly enhance plant growth and defense system (Khan *et al.,* 2009). In addition, seaweed extracts comprise different types of strong antioxidants including carotenoids, and also phenolic complexes such as; flavonoids, benzoic acid, phenolic acids, isoflavones, cinnamic acid and quercetin (Ali *et al.,* 2021). On the other hand, the consequence use of seaweed substances on crops is affected and depends on application type, mechanizes of use, and plant genotype (Ali *et al.,* 2016). These extracts are natural bioactive compounds; they can be used as liquid extracts for foliar and soil applications, or in granular form as soil improvers and fertilizer (Thirumaran *et al*., 2009).

Also, SEs have been shown to affect plant metabolism, and recent gene expression analyses have provided preliminary insights into some metabolic pathways, observed an increase in phenolic and flavonoid contents in spinach treated with *Ascophyllum nodosum* extract. This was due to the increase in the number of transcripts of key enzymes involved in nitrogen metabolism, antioxidative capacity (glutathione reductase), and glycine betaine synthesis (Kocira *et al.,* 2018). In addition, SEs itself comprise amount of macro and micronutrients that affected on improving the nutrients by enhancing nutrients translocation and metabolism in the plants (Rasouli *et al.,* 2022).

In fact, that whole SEs components can act together so as to have general positive response in the plant system, consequently each component might work independently or interactively on many metabolic networks (Malaguti *et al.,* 2002). Therefore, improve the nutrient uptake and nutrient use efficiency for both macroelements and microelements was found by the applications of these natural substances (Battacharyya *et al.,* 2015 and Colla *et al.,* 2015b). In addition, (Du Jardin, 2015) reported that morphological root characteristics, including density and root length, number of root hairs and hence their surface area has been mostly associated increase nutrient use efficiency . The biostimulant-mediated positive effects on photosynthesis, plant nutrition and secondary metabolism can improve the quality of vegetables (Battacharyya *et al.,* 2015and Colla *et al.,* 2015b).

## 2.4.1.4 Moringa leaf extract (MLE)

Moringa is a miracle plant, among the medicinal, agricultural and horticultural crops, belongs to the family *Moringaceae* (Abdrani, *et al.,* 2018). According to (Tahir *et al.,*2020), 13 species of various moringa plants which includes different sizes from small plants (*M. oleifera*) to huge trees (*M. stenopetala)* (1-15 m height). It grows quickly in tropical and subtropical regions around worldwide, but parts of Africa, Asia (sub-Himalayan tracts of India, Pakistan and Bangladesh) may be an origin place (Fahey, 2005). Moringa is rich in allelochemicals (amino acids,Threonine, methionine and phenylalanine)**,** fatty acids (Palmitic, oleic and linoleic), phenols (Gallic acid, p-coumaric acid and ferulic acid) flavonoids (Catechin, quercetin, kaempferol and niazimicin) and other bioactive compounds, vitamins (B, A, C, D and K), zeatin and essential macro elements (Potassium, magnesium and phosphorus) and microelements (Iron and zinc) (Rady and Mohamed, 2015). Tahir *et al.,* (2022)b was reported that these elements can improved growth and developed plant, tolerance to biotic and abiotic stresses and resistance against pests and diseases.

As mention before, that MLE is rich in various nutritional elements and bioactive compounds and researcher was also concluded that *Moringa* extracted can be used as a non-microbial plant bio-stimulants for promote seed germination, seedling growth and over all plant growth while low concentrations are stimulatory to test plants (Tahir *et al.,* 2020). Therefore, the impact of moringa leaf extract will be conducted on improving plant growth, yield and fruit quality of vegetable crops. Macro- and micronutrients, such as N, P, K, Ca, B, Mg, Cu, Zn, Mn, Na and Fe were observed from Moringa leaves (Yasmeen *et al.*, 2014). Therefore, those mineral nutrients from moringa leaf can be used as a natural fertilizer to improve horticultural plants and thus to decrease the levels of chemical fertilizers applied (Zulfiqar *et al.*, 2020).

**2.4.1.5 Effect of MLE on improving plant growth, yield and fruit quality of vegetable crops**

In all higher plans, Plant-derived biostimulants PDBs as natural products are presented that can be used in high value production systems in agriculture and also it can sustainability increased yield and quality, particularly under low input conditions (Kurepin *et al.*, 2014). In recent years, among plant-derived biostimulants, (MLEs) have an important role because of their significant impact on crop yield, on the other hand, MLE used as alternative to artificial chemical fertilizer in organic farming (Yasmeen, *et al.*, 2014). In addition, nutrient use efficiency, (pre- and post-harvest) characterized, quality and quantity crop and resistance against to abiotic stresses can be improved by applying moringa leaf extract. (Zulfiqar *et al.*, 2020).

Numerous researchers were verified the role and effect of MLE on growth and fruit quality on vegetable plants. Iqbal *et al.*, (2013) reported that water extracts of moringa leaf significantly impact the yield when it applied in low concentration. This research indicated that MLE juice contain substances that promoted the vegetative growth and grain yield of many crops by 25-30%. In addition, a study found that MLE most likely responsible for the growth and yield because of mineral nutrient elements, growth enhancing hormones that have been found in MLE. Analysis showed that MLEs contain gibberellins, IAA and ABA (the latter in very small quantities) (Rady and Mohamed, 2015).

Researchers demonstrated that foliar and soil application include cytokinins (*trans*-zeatin) synthetic by PGRs and MLE can significantly affected quality and quantity of cherry tomatoes under greenhouse condition, as compared to untreated plants (Basra, and Lovatt, 2016) In addition, Abdalla, (2013) reported that chlorophyll contents and photosynthesis rate can increased by applied MLEs under normal and stress conditions which is cause increase yield. Researcher also demonstrated that exogenous MLEs application on rocket (*Eruca vesicaria* subsp. *sativa*) plant can improve pigments and rate of photosynthetic with stomatal conductance as compared to control plant. MLE foliar application under normal growing conditions into snap bean (*Phaseolus vulgaris*) had significant impact on growth and development, biochemical, increased chlorophyll pigments and yield quality as coper to control plant (Elzaawely *et al.*, 2017). Researcher was concluded that added (50 gL$^{-1}$) of MLE significantly increased crop quality of fruit weight and diameter, enhanced vine length and development of total soluble solids, titratable acidity and ascorbic acid content of cucumber plant (Ullah, *et al.*, 2019).

**2.4.1.6 Effect of MLE on nutrients and phytochemical contents of the plant.**

Aslam *et al.* (2016) reported that application of MLE may able to adjust both primary and secondary metabolism, consequence increased concentration of antioxidant compounds including; concentration of phenolic antioxidants, total soluble proteins and other bioactive compounds can increase in spinach plants by apply of MLE and also supplemented with synthetic growth regulators. Mineral nutrients and fibers from fresh fruits and vegetables are particularly encouraged for consumer as well as properties of natural phytochemical content regarding the health benefiting (Zulfiqar *et al.*, 2020).

Nasir, *et al.* (2016) demonstrated that foliar application of MLE individually and combination with Zn and K, significantly improved leaf N, P, K, Ca, Mg and Zn contents in a field study with 'Kinnow' mandarin, results confirmed that increased leaf nutrient content may be due to MLE minerals. In addition, spray of moringa leaf and twig extract on rocket (*Eruca vesicaria* subsp. sativa) plants, can increased nutrient content including N, P, K, Ca, Mg, and Fe consequently improved plant growth (Abdalla, 2015). In addition, Yasmeen *et al.*, (2014). Confirmed that Moringa leaves also contain macro- and micronutrients, such as N, P, K, Ca, B, Mg, Cu, Zn, Mn, Na and Fe.

As mention before, moringa considered as a good source for phytohormones that have a positive role is stimulating growth and productivity of plant. Elzaawely *et al.*, (2017) concluded that the MLE on snap bean plants had significant impact on plant growth, and also increased phytochemical and growth hormone content. In addition, researcher reported that using concentration 1:20 of MLE can increased ABA content while used (1:30) concentrations causes decrease the rate of ABA they, also concluded that increase in endogenous level of gibberellin (GA7) as enhanced photosynthetic activity broader leaves consequently increased pod yield. Ahmed, *et al* (2020) concluded that foliar application of (MLE) had the ability to promote the vegetative growth and physiological traits of cucumber (*Cucumis sativus* L.) leading to the positive enhancement in the yield and nutritional quality of cucumber fruits. This promotion of the physiological traits of cucumber plants was accompanied by an increase in the endogenous hormone levels and enzymes activity. This study proved and support the assumption that MLE might be utilized as effective natural and most importantly safe biostimulants to decrease chemical fertilizer during organically production and consequently substitution of hazard chemicals by natural ones in the production of high-quality vegetables and fruits.

**2.4.2 Microbial plant biostimulants.**

Microbial plant biostimulants are inventive technologies able to enhance the nutritional values and ensure agricultural productivity with overcoming the negative impacts derived from environmental changes (Castiglione *et al.*, 2021). The microbial biostimulants is living microorganism (or mix of microorganisms) which is added to seed, plant surface, colonize the rhizosphere, and also it can enhance plant growth via obtainability of primary nutrients to the host plant (Barman *et al.,* 2017). Bacteria, fungi, and other microbes a vital and asymmetrically distributed micro-organism were presented everywhere in the world with different number of concentrations, but appear in more concentration near the root and perform very important works (Akram, *et al.,* 2020). Therefore, the essential unit of soil system is microorganisms which are play a significant role in rhizosphere activities, it can implicate the soil promising for nutrient mineralization and sustainability to crop production (Mahanty *et al.,* 2017). In addition, in the rhizosphere zone high density and diversity of microbial is indicated by the variety, morphology architectures of roots, and displayed of organic nutrients quantity, the biotic and abiotic factors community, plants themselves (Enagbonma and Babalola, 2019). The microbial diversity of the soil rhizosphere is indicated by the diversity and quantity of organic nutrients exuded, root system architecture, branching order of root (Pervaiz *et al.,* 2020). Therefore, Bio-fertilizers are carrier-based products having living strains of efficient microorganism that accelerates nutrient uptake of crop by changing rhizosphere developments when added through soil and seed (Singh *et al.,* 2020).

Furthermore, one of the important advantages of microbial biostimulants is that can be used individually or in combination with other fertilizers (Boraste *et al.,*2009). Other advantages, they are economical and natural source of macronutrients and micronutrients, production consequently, plant growth promoting hormones, promoting soil health and justifying the harmful impact of chemical fertilizers. In addition, one of the benefits of MBS is that plays significant role in producing nutrient and make them available to the plant in the soil Barman *et al.,* (2017) that produce nutrients available logically plentiful in soil and atmosphere to crops. Various researcher, were confirmed that it is cheap, ecofriendly, free from any ecosystem risks (Sahoo *et al.,* 2014 and Borkar, 2015).

On the other side, chemical fertilizers and pesticides are expected to providing by presenting biofertilizers (Subashini *et al.,* 2007). Furthermore, biofertilizers able to adapt significant nutritional elements from unobtainable to obtainable via organic processes (Vessey, 2003). Natural nutrient cycle could regenerate by microorganisms in biofertilizers result, it can maintain optimum nutrient level and organic matter in soil, consequently developed good plant and keeping soil fertility (Singh *et al.,* 2011; Sinha *et al.,* 2014 and Shelat *et al.,* 2017). Besides accessing nutrients biofertilizers deliver growth-promoting factors to plants via excretion of different vitamins, phytohormones (Revillas *et al.,* 2000 and Abd El-Fattah *et al.,* 2013). They by successfully facilate composting and controlling attack of soil borne diseases and pests (Board, 2004 and Sinha *et al.,* 2014).

# CHAPTER THREE

# MATERIALS AND METHODS

## 3.1 The Experiment Site

The experiment was conducted in a greenhouse during the period of May 3, 2021 to August 5, 2021 at the research farm belongs to the Horticulture Department, College of Agricultural Engineering Sciences, University of Sulaimani, Sulaimani, Iraq (latitude: 35º 32' N, longitude: 45º 21' E, altitude 730 meter above sea level). The greenhouse dimensions were (40 m length, 9 m width, and 3.2 m height) which it was covered by polyethylene plastic film with 200μm thickness.

## 3.2 Soil Preparation, Experiment Setup and Soil Analysis

The soil of the greenhouse was plowed, then smoothed and leveled using the rotivator machine. A distance of 100cm was left from both side inside the greenhouse, and the remaining distance (700 cm) was divided into four terraces and five gangways of 85 and 72 cm width, respectively. Each terrace consists of two cultivation lines; space between the plants within a line was 40 cm. The experimental units were separated with a length of 260 cm; and 120 cm was left between two units. As a result, the area of each experimental unit was 4.082 m² (260 × 157 cm). Furthermore, each experimental unit contained 12 plants resulting in a plant density (2.93 plants m$^{-2}$) (Appendix 1). During the field preparation, the cattle and sheep manure at a rate of (4 kg per unit) and poultry manure at a rate of (1.850 kg per unit) were homogeneously mixed with the soil of the experimental units, after which the drip irrigation system was installed. Thus, the greenhouse was ready for cultivation.

After soil preparation, soil samples were taken from the depth of (0-30 cm, 12 sub samples) to determine baseline soil properties. The samples were air dried and passed through a 2 mm sieve prior to analysis. The analyses were conducted at Bazian laboratory, which is a private lab for chemical, physical and biological analysis of soil, water and plant in Sulaimani city. The results of some physical and chemical analysis of the soil are presented in (Table 3.1).

Table 3. 1: Some of the physical and chemical characteristics of the experiment soil.

| Soil characteristics | Values | Unites |
|---|---|---|
| pH | 7.17 | ----- |
| EC | 1.58 | ds m$^{-1}$ |
| Organic matter | 33.0 | g kg$^{-1}$ |
| CaCO$_3$ | 140.00 | g kg$^{-1}$ |
| Total N | 0.77 | g kg$^{-1}$ |
| Available P | 5.88 | mg kg$^{-1}$ |

| Available K | 38.40 | g kg$^{-1}$ |
| --- | --- | --- |
| Available Zn | 0.70 | mg kg$^{-1}$ |
| Available Fe | 6.00 | mg kg$^{-1}$ |
| Sand | 505.00 | g kg$^{-1}$ |
| Silt | 293.80 | g kg$^{-1}$ |
| Clay | 201.20 | g kg$^{-1}$ |
| Texture | Loam | ----- |

## 3.3 Plant Materials

The cucumber seeds (Sayff-F1), which is an F1 hybrid and produced by Nunhems®, was used in this study. This hybrid is widely cultivated under protected conditions by farmers in Kurdistan Region because of its high yield potential and appealing fruit shape by consumers. According to Nunhems ®, general characteristics of the Sayff-F1 cucumber are shown in (Appendix 2).

## 3.4 Experimental Treatments

The experiment was designed to study the effect of some biostimulants individually and their combinations on growth, yield and fruit quality of cucumber (Sayff- F1) under greenhouse conditions. In addition, to compare the results with results of chemical fertilizer application or conventional fertilizers. The treatments were as the following:

- *T1:* Control: without adding any kinds of biostimulants
- *T2:* Recommended dose of chemical fertilizers (RDCF)
- *T3:* Non-microbial biostimulant of seaweed extracts (SE)
- *T4:* Plant-based biostimulant of moringa leaf extract (MLE)
- *T5:* Bacterial-based biostimulant of Fulzym- Plus (FP)
- *T6:* SE + MLE
- *T7:* SE + FP
- *T8:* MLE + FP
- *T9:* SE + MLE + FP

## 3.5 Seedling Preparation and Transplanting

The cucumber seeds were sown in May 3, 2021 in the seedling trays, which filled with a mixture of sterilized peat-moss and perlite. The seeds were sown under glasshouse conditions, which were maintained at 25/18 ± 2°C Day/night temperature, 14/10 h light/dark photoperiod, and a relative humidity of 65 ± 10%. During the seed sowing, half of the seeds were treated with the microbial plant biostimulats of FP at a rate of (2 kg ton$^{-1}$ seed) by seed dressing method. After 15 days of seed sowing, the seedlings were transplanted at two true leaves stage, which were cultivated on the terraces in zigzag pattern. Before transplanting, seedlings that inoculated with FP were treated again with the same inoculum by root dipping method at a rate of (10g L$^{-1}$). In order to study the effect of the treatments on the plant root system in depth, one seedling from each experimental units were cultivated in a black polyethylene pot (26 cm diameter, 26 cm height, and 13.5 kg of soil capacity) to facilitate getting the entire roots system easily and safely.

## 3.6 Treatments Application

### 3.6.1 Application of T2 (RDCF)

This treatment included the application of macro and micronutrients and applied in two methods: soil application and foliar application. It was implemented after a week of transplanting (Table 3.2).

**Table 3. 2 Applied recommended dose of chemical fertilizers (RDCF)**

| Weeks after transplanting | Chemical fertilizers types (Soil application) | Dosages (g plant$^{-1}$) | Chemical fertilizers types (Foliar application) | Dosages (g L$^{-1}$) |
|---|---|---|---|---|
| 2$^{nd}$ | NPK 10:40:10 | 2.0 | NPK (12:48:8) | 2.0 |
| 3$^{rd}$ | NPK 10:40:10 | 2.0 | NPK 20:20:20 | 2.0 |
| 4$^{th}$ | NPK 20:20:20 | 2.5 | Excellent * | 1.0 |
| 5$^{th}$ | NPK 20:20:20 | 2.5 | NPK 20:20:20 | 2.5 |
| 6$^{th}$ | Calmag+Zn | 3.0 | Excellent | 1.0 |
| 7$^{th}$ | NPK 12:10:36 | 5.0 | NPK 9:15:30 | 2.5 |
| 8$^{th}$ | Calmag+Zn | 3.0 | NPK 20:20:20 | 2.5 |
| 9$^{th}$ | NPK 12:10:36 | 5.0 | NPK 9:15:30 | 2.5 |

* The chemical composition of the Excellent (Napnnutriscience Co., Ltd, Thailand) foliar fertilizer is shown in (Appendix 3)

### 3.6.2 Application of T3 (SE):

The commercial product (Alga 600), which is the soluble seaweed extract powder and produced by LEILI $^®$ Company, Beijing, China (Appendix 4), was applied four times during the growing season at a rate of (2 kg ha$^{-1}$) by fertigation method. The first application was conducted after 10 days of transplanting and the others at two weeks intervals.

### 3.6.3 Application of T4 (MLE):

In order to preparation of the MLE, fresh leaves of moringa tress (*Moringa oleifera*) were collected in Sulaimani city. Aqueous extract of the collected leaves was prepared according to Ahmed *et al.,* (2020). Briefly, 30 g of the fresh leaves was mixed with 300 mL of distilled water using household blender for 15 minutes. The homogenate of the leaves was filtered through two layers of Miracloth, and then the volume of the filtrate was adjusted to the ratio of 1:10 (v/v) by distilled water. The extract was diluted with distilled water at the ratio of 1:20 (v/v). After extraction process, the extract was sprayed immediately after preparation three times during the growing season which was started after 3 weeks from transplanting and the others at 15 days intervals.

## 3.6.4 Application of T5 (FP):

FP is a bacterial-based biostimulant produced by OMRI® Company, USA. This product contains beneficial bacteria of *Bacillus subtilis* and *Pseudomonas putida* ($2 \times 10^{10}$ per gram); and enzymatic systems of protease, amylase, chitinase and lipase; as well as plant growth promoters of gibberellins and cytokinins (Appendix 5). This treatment applied by the following methods:

- seed dressing before sowing at a rate of (2 kg ton$^{-1}$ seed),
- seedlings root dipping in the bacterial suspension before transplanting at a rate of (10 g L$^{-1}$), and
- fertigation method after transplanting at a rate of (1Kg ha$^{-1}$), which applied once every two weeks from transplanting for four times.

## 3.7 Growth Conditions

During the experimental period, the air temperature and relative humidity inside the greenhouse were measured by using data logger device (Model: Perfect-Prime TH0160) which the data was recorded hourly. The device was placed in the middle of the greenhouse at a high of 1.5m above the soil surface. Furthermore, the meteorological data (maximum, minimum and average of temperature and relative humidity) during the study period inside and outside the greenhouse are summarized in (Table 3.3).

**Table 3. 3: Monthly maximum, minimum and average of temperature and relative humidity outside and inside the greenhouse after transplanting until the end of the season**

| Months | Outside the greenhouse* | | | | | | Inside the greenhouse | | | | | |
|---|---|---|---|---|---|---|---|---|---|---|---|---|
| | Temp. ˚C | | | RH (%) | | | Temp. ˚C | | | RH (%) | | |
| | Max. | Min. | Ave. | Max. | Min. | Ave. | Max. | Min. | Ave. | Max. | Min. | Ave. |
| From 17, May | 35.4 | 20.3 | 27.9 | 47.9 | 18.5 | 33.2 | 38.8 | 23.9 | 31.4 | 68.5 | 38.9 | 48.3 |
| June | 40.0 | 22.6 | 31.3 | 39.9 | 15.1 | 27.5 | 43.6 | 24.3 | 33.3 | 59.3 | 33.6 | 42.9 |
| July | 43.2 | 27.4 | 35.3 | 39.3 | 18.0 | 28.6 | 52.2 | 28.3 | 34.4 | 48.8 | 28.4 | 41.1 |
| August | 43.0 | 26.6 | 34.8 | 39.0 | 16.8 | 27.9 | 51.4 | 24.1 | 35.2 | 45.9 | 26.8 | 38.3 |

*Data obtained from the agro-meteorological station belongs to the college of agricultural engineering sciences, in Bakrajo, Sulaimani; which is two kilometers far away from the experiment site.

## 3.8 Agronomic Practices

All the common agronomic practices for cucumber production inside the greenhouses such as irrigation, weed control, ventilation, thinning and pruning were carried out uniformly during the growing season for all treatments. Plants were pruned to a single stem and supported by strings. In addition, lower leaves were removed to a height of 50 cm from the soil surface. Anti-insect net with 50 meshes was used for vents to prevent entering harmful insects inside the greenhouse.

## 3.9 Experimental Design and Data Analysis

The experiment was laid out in a randomized complete block design (Simple- RCBD) with three replications, each replication consisted of nine experimental units and the treatments were distributed randomly within each replicate, as shown in the experiment layout (Appendix 1) The analysis of variance (One -way ANOVA), and Duncan's new multiple range test at P≤ 0.05 were implemented using XLSTAT software. Mean squares (MSs) of the analysis of variance for all the studied characteristics are shown in (Appendices 6, 7, 8, 9, 10 and 11).

## 3.10 Measurements

### 3.10.1 Root measurements

At the end of the season, the root system of the representative plants, which were planted in the black plastic pots, was extracted and cleaned manually using tap water and the following characteristics were registered:

#### 3.10.1.1 Number of the lateral roots per plant

The number of the lateral roots per plant was measured by counting the branches of the root system.

#### 3.10.1.2 Lateral root length (cm)

Lateral root length was measured from the crown zone to its farthest point by using the metric tape.

#### 3.10.1.3 Lateral root diameter (mm)

Lateral root diameter was measured from the middle of the roots by using digital caliper, and then the mean was calculated.

#### 3.10.1.4 Root system dry weight (g)

Root system dry weight was determined after drying the root samples in the forced-air oven (Model: LOD-250N, LabTech®, Korea) at 68°C for about 72 hours, until the weight was stable, and then the dried samples were weighted by a digital scale.

**3.10.2 Vegetative growth measurements**

At the end of the season, six plants for each experimental unit were selected randomly to measure some of the vegetative growth characteristics, including:

**3.10.2.1 Plant shoots height (cm)**

Plant shoots height was measured from the base of the stem to the top of the plants using metric tape, and then the averages were calculated.

**3.10.2.2 Number of nodes per plant**

Number of nodes per plant was measured by counting the number of nodes on the main stem of the selected plants and then the averages were calculated.

**3.10.2.3 Stem diameter (mm)**

Stem diameter of the selected plants was recorded using digital caliper at three positions for each plant, (50, 100 and 150 cm) above the soil surface, and then the averages were calculated.

**3.10.2.4 Plant leaf area (dm²)**

Plant leaf area was calculated by measuring the area of six randomly selected leaves in each experimental unit by using Digimizer program (v.4.5.2®) (Rasul 2023), and then the average of single leaf area was multiplied by the total number of the leaves per plant to calculate the plant leaf area

**3.10.2.5 Leaf dry matter (%)**

Leaf samples were taken from the selected plants in each experimental unit. These samples, after measuring their fresh weights, were placed in an electric oven with vacuum at 68 °C for 48 hours until the dry weight stabilized. The percentage of dry matter was calculated as follows:

$$\text{Leaf dry matter (\%)} = \frac{\text{Dry weight of the sample (g)}}{\text{Fresh weight of the sample (g)}} \times 100$$

**3.10.2.6 Leaf chlorophyll pigment intensity (SPAD)**

Leaf chlorophyll intensity was measured by using portable chlorophyll meter (SPAD-502; Minolta, Osaka, Japan).

### 3.10.3 Leaf nutrients analysis

The fourth leaf from the meristem point of six randomly selected plants of each experimental unit was collected to determinate the concentrations of some macro and micronutrients (Maboko *et al.*, 2013). The selected leaves were oven dried with the vacuum at 68˚C for 48 hours and then grinded with grinder. The samples were digested by taking 200 mg of the grinded leaf samples and digested with concentrated sulfuric and perchloric acids with the ratio of (5: 3) according to Creeser and Parsons (1979). After the digestion process the following nutrients were estimated; and the calculations of all nutrient concentrations were expressed on a dry mass basis. The analysis was conducted at Central Laboratory for soil water analysis, College of Agriculture engineering Sciences, University of Baghdad.

### 3.10.3.1 Nitrogen (mg/gm)

Nitrogen (N) was estimated by evaporation and distillation process with Micro- Kjeldahl according to the method proposed by Jackson (1958).

### 3.10.3.2 Phosphorous (mg/gm)

Phosphorous (P) was estimated by Spectrophotometer at 882 nm wavelength according to Olsen and Sommers (1982).

### 3.10.3.3 Potassium (mg/gm)

(K) was determined by using Flame photometer according to (Page *et al.,* 1982).

### 3.10.3.4 Micro nutrients of Iron and Zinc (mg Kg$^{-1}$)

Iron (Fe) and Zinc (Zn) were determined by using Atomic Absorption Spectrophotometer (Jaiswal, 2004).

### 3.10.4 Reproductive growth and yield component characteristics

The means were calculated for six plants from each experimental unit for the following reproductive and yield component characteristics:

### 3.10.4.1 Number of flowers per plant

The number of flowers per plant was measured by counting the flowers on the selected plants during the study period and then the mean was calculated.

### 3.10.4.2 Number of aborted flowers per plant

The number of aborted flowers per plant was measured by counting the number of the aborted flowers and then the average per plant was calculated.

### 3.10.4.3 Fruit set (%)

Fruit set was calculated by dividing the mean of number of fruits per plant by the total number of flowers per plant as following:

$$\text{Fruit set (\%)} = \frac{\text{Number of fruits per plant}}{\text{Number of flowers per plant}} \times 100$$

### 3.10.4.4 Number of fruits per plant

The average number of fruits per plant was measured by counting the fruits from the beginning of the harvesting (June 15, 2021) to the end of the season (August 5, 2021).

### 3.10.4.5 Fruit weight (g)

The average weight of a single fruit was calculated by dividing the total yield of the six plants which selected in each experimental unit by the total number of fruits for the same plants.

### 3.10.4.6 Early yield (Kg plant$^{-1}$)

The first three harvests were considered as the early yield, during the period from (June 15, 2021) to (June 22, 2021).

### 3.10.4.7 Plant yield (Kg plant$^{-1}$)

The fruits were harvested every two to four days. During the production period, the total weight of the fruits from the selected plants in each experimental unit was measured and then the average was calculated to determine the yield per plant in kilograms.

### 3.10.5 Fruit quality characteristics

To illustrate the import of the studied factors on the fruit quality the following fruit characteristics were measure it.

### 3.10.5.1 Fruit length (cm)

Average fruit length was measured by taking 10 fruits randomly from each experimental unit for two harvesting times by using ordinary ruler.

### 3.10.5.2 Fruit diameter (mm)

The average diameter of the fruits was measured for the same fruits that were used to measure their lengths by using digital caliper.

### 3.10.5.3 Fruit dry matter (%)

Slices of 10 fruits were taken from each experimental unit, and the samples immediately were placed in a forced air oven with vacuums at a temperature of 70˚C for 72 hours, until the weight was stable, and then the percentage of dry matter in the fruits was estimated according to the following equation:

$$\text{Fruit dry matter (\%)} = \frac{\text{Weight of the slices after drying}}{\text{Weight of the slices before drying}} \times 100$$

### 3.10.5.4 Total soluble solid (TSS)

The TSS of the fruits was determined using a portable digital refractometer (Model: PAL-1, Atago, Tokyo Tech., Japan) (Rasul *et al.,* 2022). The refractometer was calibrated first with distilled water. Following that, two drops of the fruit juice were placed on the device's sensor, and the TSS was read in °Brix.

### 3.10.6 Fruit non-enzymatic antioxidants analysis

For determining some of the non-enzymatic antioxidants in the fruits; such as total phenolic content (TPC) and total flavonoid content (TFC) the method of Michiels *et al* (2012) and Tahir *et al.* (2023) was used for preparing the sample extracts as follows:

Fruit samples were taken randomly from each experimental unit, and they were snap frozen by using liquid nitrogen. Frozen samples were ground to fine powder using pestle and mortar. 0.4 g of ground powder were taken and putted in 2 ml Eppendorf tube, then 1 ml of 60% methanol was added into the tubes and mixed gently. The mixture was incubated for 16 hours and then centrifuged at 10000 rpm for 15 minutes at 4°C. The supernatants layer (extracts) was transferred into new Eppendorf tubes and they were kept in refrigerator at 4°C as a crude extracted solution for non-enzymatic antioxidants analysis.

### 3.10.6.1 Total phenolic content (TPC) determination (µg GAE $g^{-1}$ FW)

The Folin-Ciocalteu method was applied to determine TPC as described by (Tahir *et al.* 2022b; Faraj 2023) with some modifications. Briefly, 600 µL of each extract was mixed with 2.2 ml of Folin-Ciocalteu reagent. After 7 min of incubation, 1.8 mL of sodium carbonate solution (10%) was added and left in the dark at 38˚C for 50 minutes. Regarding the blank, the same previous step was repeated but only 600 µL of distilled water was used instead of the sample extracts. The absorbance of the reaction mixture was registered at 750 nm by spectrophotometer (Model: UV-160, Shimadzu, Japan).

The standard solution was prepared by dissolving 9 mg of gallic acid in 9 mL of methanol to achieve a final concentration of 1 mg mL$^{-1}$. Series of dilutions of gallic acid (0, 50, 100, 150, 200, 250, and 300μg mL$^{-1}$) had been used to produce a standard curve and linear regression between the absorbance values at 750 nm and the gallic acid concentrations was found. The TPC was calculated and expressed as micrograms of gallic acid equivalent (GAE) per gram of fresh weight of the fruit samples (μg GAE g$^{-1}$ FW) by using the following equation:

$$\text{TPC (μg GAE g}-1\text{ FW)} = \frac{V}{W} \times C$$

Where:

V: is the volume of the extracts (mL),

W: is the fresh weight of the samples (g), and

C: is the concentration of gallic acid determined from the standard curve.

### 3.10.6.2 Total flavonoid content (TFC) (μg QE g$^{-1}$ FW)

TFC was determined according to the method described by Rigane *et al.* (2017). An aliquot of 600 μL of each fruit sample extracts was added to the mixture of 0.9 ml methanol (80%), 0.3 ml of 2% aluminum chloride (AlCl$_3$), 0.07 ml of (1M) potassium acetate (CH$_3$COOK), and 1.7 ml of deionized water. The blank contained the same amount of chemicals with 600 μl of distilled water instead of the fruit extracts. The mixture was incubated at room temperature for 30 minutes. Thereafter, the absorbance of the reaction mixture was recorded at 415 nm using spectrophotometer.

The standard solution was prepared by dissolving 9 mg of quercetin in 9 mL of methanol to achieve a final concentration of 1 mg mL$^{-1}$. Sequence dilutions of the quercetin solution (0, 2.5, 5, 10, 20, 40, and 80 μg mL$^{-1}$) was used to produce a standard curve. Then, a linear association was observed between the absorbance values at 415nm and the quercetin concentrations. The TFC was calculated and the results were expressed as micrograms of quercetin equivalents (QE) per gram of the fresh fruit samples (μg QE g$^{-1}$ FW) by using the following equation:

$$\text{TFC (μg QE g}-1\text{ FW)} = \frac{V}{W} \times C$$

Where:

V: is the volume of the extracts (mL),

W: is the fresh weight of the samples (g), and

C: is the concentration of quercetin collected from the standard curve.

# CHAPTER FOUR

# RESULTS AND DISCUSSION

## 4.1 Effect of Biostimulants on Cucumber Root Growth Characteristics

### 4.1.1 Number of lateral roots per plant

The influence of non-microbial biostimulants (SE and MLE) as well as bacterial- based biostimulants of Fulzym- plus (FP) on number of lateral roots per cucumber plant is shown in (Table 4.1). The showed Results show that the application of different biostimulants, individually and their combinations, had a significant impact on improving this trait. However, the highest number (9.33) was recorded by the application of recommended dose of chemical fertilizer (RDCF). While, this treatment was not differed substantially with (SE + FP + MLE) which was registered 9.00 lateral roots per plant. In addition, cucumber plants grown without adding any kinds of biostimulants (Control) registered the minimum number of lateral roots per plant which was recorded (3.67). In comparison to control, the application of bistimulants of (SE), (MLE) and (FP) caused significant increase of lateral roots number per plant by 99.7, 72.5 and 54.5%, respectively. Also, the combination effects of these biostimulants in the treatments of (SE + MLE), (SE + FP), (MLE + FP) and (SE + MLE + FP) significantly increased this trait by 81.7, 72.5, 90.7 and 145.2%, respectively.

### 4.1.2 Lateral root length (cm)

Average lateral root length was affected significantly by applying recommended dose of chemical fertilizers as well as different biostimulants, either individually or in combinations (Table 4.1). The longest lateral root (70.00 cm) was recorded by the application of recommended dose of chemical fertilizer (RDCF), which it was not differ significantly with the treatments of (SE + MLE) and (SE + MLE + FP) which were registered 63.33 and 68.33 cm, respectively. However, the shortest lateral root was observed by the control (36.33 cm). If compared to control, increasing percentage of lateral root length reached 44.0% for (SE), 53.7% for (MLE), 53.2% for (FP), 74.3% for (SE + MLE), 52.3% for (SE + FP), 68.8% for (MLE + FP) and 88.1% for (SE + MLE + FP).

### 4.1.3 Lateral root diameter (mm)

Results in Table (4.1) illustrate that lateral root diameter was affected significantly by using non- microbial and bacterial- based biostimulants. Although, the largest diameter of lateral roots (3.03 mm) was obtained by adding chemical fertilizer (RDCF) but it was statistically similar with the treatments of (FP) and (SE +

FP + MLE) which were recorded 2.85 and 2.98 mm, respectively. In addition, the lowest diameter of lateral roots was recorded by control (1.21 mm). Additionally, the biostimulant treatments of (SE), (MLE) and (FP) significantly improved this character by 69.4, 81.0, and 135.5%, respectively. Also, increasing percentages for the combination treatments of (SE + MLE), (SE + FP), (MLE + FP) and (SE + MLE + FP) were 68.6, 66.9, 90.1 and 146.3% respectively, in comparison with control.

### 4.1.4 Root system dry weight (g)

Data in Table (4.1) proved that the application of non-microbial biostimulants (SE and MLE) as well as bacterial- based biostimulants of Fulzym- plus (FP), separately and in combinations, had a significant impact on enhancing cucumber roots system dry weight. However, the highest dry weight of the roots (24.10 g) was observed by the application of chemical fertilizers (RDCF), but it was not statistically superior to the treatments that consists of the combinations of the biostimulants (SE + MLE), (SE + FP), (MLE + FP) and (SE + MLE + FP) which were recorded 19.91, 21.67, 21.01 and 23.67 g respectively. In addition, cucumber plants grown without adding any kinds of biostimulants (Control) has the lightest dry weight of the roots (9.84 g). Therefore, if compared to control, the application of biostimulants of (SE), (MLE) and (FP) caused significant increase of root dry weight by 78.3, 61.9 and 91.4%, respectively. Also, the combination effects of these biostimulants in the treatments of (SE + MLE), (SE + FP), (MLE + FP) and (SE + MLE + FP) significantly increased this trait by 102.3, 120.2, 113.5 and 140.5%, respectively.

**Table 4. 1: Effect of biostimulants on cucumber root growth characteristics**

| Treatments | No. of lateral roots per plant | Lateral root length (cm) | Lateral root diameter(mm) | Root system dry weight(g) |
|---|---|---|---|---|
| Ctrl. | 3.67  d | 36.33  e | 1.21  c | 9.84  d |
| RDCF | 9.33  a | 70.00  a | 3.03  a | 24.10  a |
| SE | 7.33  b | 52.33  d | 2.05  b | 17.54  bc |
| MLE | 6.33  bc | 55.83  cd | 2.19  b | 15.93  c |
| FP | 5.67  c | 55.67  cd | 2.85  a | 18.83  bc |
| SE + MLE | 6.67  bc | 63.33  abc | 2.04  b | 19.91  abc |
| SE + FP | 6.33  bc | 55.33  cd | 2.02  b | 21.67  ab |
| MLE + FP | 7.00 bc | 61.33  bc | 2.30  b | 21.01  ab |
| SE + MLE + FP | 9.00  a | 68.33  ab | 2.98  a | 23.67  a |

Ctrl: control; RDCF: recommended dose of chemical fertilizer; SE: seaweed extracts; FP: microbial- based biostimulant of Fulzume-plus; MLE: moringa leaf extract.

Different letters in the same column indicate significant differences between means according to Duncan's new multiple range test at P≤ 0.05

The primary functions of the root are to support the aboveground biomass and to uptake the water and nutrients from the soil to the other part of the plant in order to enhance plant growth and development (Grover *et al.,* 2021). Therefore, plant survival very much relies on root system architecture. Modification of root characteristics could contribute to enhancements of desirable agronomic traits such as vegetative growth, flowering growth, plant yield and fruit quality (Siddique *et al.*, 2015). In addition, improving root system characteristics are important in view of current challenges such as sustainable and/ or organic crops production (Leitner *et al.*, 2014). The results of our study showed that the application of different biostimulants (SE, MLE and FP), individually and their combinations, had a significant impact on improving number of lateral roots per plant, lateral root length and diameter as well as root system dry weight if compared to untreated plants (control). Furthermore, using the three biostimulants together in the treatment of (SE + MLE + FP) did not different significantly with (RDCF) in all studied root system characteristics (Table 4.1).

SE contributed to the enhancement of the studied root traits due to the effects of phytohormones and growth regulatory substances, including Indole-3-acetic acid (IAA) and cytokinins, in which present in the SE and induced the biosynthesis of these hormones by the treated plants (Ali *et al.*, 2019). IAA is responsible for a wide range of growth processes such as cell division, vascular growth, elongation and differentiation of roots. Cytokinins account for root: shoot ratio, nutrient mobilization and delay in senescence (Zhang and Ervin, 2004). For these reasons, the SE boosted cucumber lateral roots number per plant in comparison with untreated plant by 99.7%, lateral root length by 44.0% and lateral root diameter by 69.4% and consequently root system dry weight by 78.3%.

Furthermore, MLE characterized by a relatively high contains of macro and micronutrients, amino acids and vitamins; in addition to secondary metabolites including antioxidants and phytohormones (Gopalakrishnan *et al.*, 2016; Brazales-Cevallos *et al.*, 2022). So, in the current study its foliar application had a positive effect on cucumber root morphology traits, with an increase of 72.5, 53.7, 81.0 and 61.9% for number of lateral roots per plant, lateral root length and diameter and root system dry weight, compared to untreated plants, respectively.

Moreover, FP contains beneficial bacteria of *Bacillus subtilis* and *Pseudomonas putida*. These bacteria are the most important group in the plant growth-promoting rhizobacteria (PGPR). PGPR colonize plant roots and they can modulate root growth and development via the production of phytohormones "especially auxins", secondary metabolites and enzymes (Mohamed and Gomaa, 2012). The most commonly observed effects are an increase of the number and length of lateral roots and root hairs. PGPR also influence plant nutrition via nitrogen fixation, phosphorus solubilization, siderophore production, and as a result, increases the dry weight of the roots. Additionally, these bacteria can modify root physiology by changing gene transcription and metabolite biosynthesis in plant cells (Vacheron *et al.*, 2013).

In addition, FP product contains enzymes of protease, amylase, chitinase and lipase; as well as plant growth promoters of gibberellins and cytokinins. Thus, in this study lateral roots number per plant, lateral root length and diameter as well as root system dry weight of cucumber plants were improved as a result of FP treatment by 54.5, 53.2, 135.5 and 91.4%, respectively. The highest averages of studied root growth characteristics, after (RDCF) treatment, were found in the triple interaction between the biostimulants (SE + MLE + FP); this may be returned to the synergistic effects of these biostimulants which resulted in better roots proliferation.

## 4.2 Effect of Biostimulants on Cucumber Vegetative Growth Characteristics

### 4.2.1 Shoot height (m)

The effect of non-microbial biostimulants (SE and MLE) as well as bacterial- based biostimulants of Fulzym- plus (FP) on cucumber shoot height is shown in (Table 4.2). No significant differences were observed between the recommended dose of chemical fertilizers (RDCF) and triple combination between the biostimulants (SE + FP + MLE); which they registered (5.30 and 4.99 m) respectively, whereas they were significantly superior over the other treatments. While, the minimum shoot height was recorded by control (3.23 m). However, the differences between the individual and dual interactions of the biostimulants were not reached the significant level, but all of them were significantly superior to the control. If compared to control, increasing percentage of the shoot height were 26.0% for (SE), 36.5% for (MLE), 25.7% for (FP), 29.1% for (SE + MLE), 33.7% for (SE + FP), 27.6% for (MLE + FP) and 54.5 for (SE + MLE + FP).

### 4.2.2 Number of nodes per plant

The number of nodes on the main stem of the cucumber plants was affected significantly by applying different biostimulants, either individually or in combinations (Table 4.2). The highest number (88.00) was observed by the application of recommended dose of chemical fertilizer (RDCF), which it was not differ significantly with different biostimulant treatments but all of them were statistically superior to control, which was recorded the minimum number (62.50).

### 4.2.3 Stem diameter (mm)

Data in Table (4.2) illustrate that the thickest stem of cucumber plants (8.92 mm) was recorded by the application of chemical fertilizers (RDCF), but control treatment had also had the lowest stem diameter which was recorded (6.45 mm). In addition, all of the biostimulant treatments were statistically superior to control except the application of (FP).

### 4.2.4 Plant leaf area (dm$^2$)

Cucumber plants that fertilized with the recommended dose of chemical fertilizers (RDCF) showed the largest leaf area (151.80 dm$^2$). On the other hand, control showed the lowest plant leaf area with an average of (94.15 dm$^2$). All the biosimulant treatments, except the application of (SE + MLE), were significantly superior to control and the percentage increase was 51.9% for (SE), 34.6% for (MLE), 37.2% for (FP), 33.1% for (SE + FP), 49.6% for (MLE + FP) and 57.2% for (SE + MLE + FP) (Table 4.2).

### 4.2.5 Leaf dry matter (%)

Results in Table (4.2) show that using non- microbial and bacterial- based biostimulants played a significant role in increasing dry matter percentage in cucumber leaves. Additionally, cucumber plants that fertilized with the recommended dose of chemical fertilizers (RDCF) gave the highest percentage of dry matter in the leaves (15.78). Also, all of the biostimulant treatments, except the application of (SE), were statistically superior to control which was recorded the lowest percentage (9.16).

### 4.2.6 Leaf chlorophyll intensity (SPAD)

As shown in (Table 4.2), no significant differences were observed between the treatments of (RDCF) and (SE + FP + MLE) in chlorophyll intensity which they recorded (47.99 and 46.13), SPAD respectively. Additionally, they were significantly superior over the other treatments. While, minimum chlorophyll intensity was recorded by control (32.48) SPAD. However, the differences between the individual and dual interactions of the biostimulants were not reached the significant level in this trait, but all of them were significantly superior to the control.

**Table 4. 2: Effect of biostimulants on cucumber vegetative growth characteristics**

| Treatments | Shoot height (m) | No. of nodes per plant | Stem diameter (mm) | Plant leaf area (dm$^2$) | Leaf dry matter (%) | Leaf chlorophyll intensity (SPAD) |
|---|---|---|---|---|---|---|
| Ctrl. | 3.23  c | 62.50  b | 6.45  b | 94.15  b | 9.16  b | 32.48  c |
| RDCF | 5.30  a | 88.00  a | 8.92  a | 151.80  a | 15.78  a | 47.99  a |
| SE | 4.07  b | 85.56  a | 8.26  a | 143.03  a | 12.15  ab | 39.03  b |
| MLE | 4.41  b | 76.28  a | 7.99  a | 126.76  a | 13.22  a | 39.87  b |
| FP | 4.06  b | 79.89  a | 7.79  ab | 129.20  a | 13.47  a | 38.00  b |
| SE + MLE | 4.17  b | 76.33  a | 8.40  a | 122.01  ab | 14.90  a | 38.73  b |
| SE + FP | 4.32  b | 75.50  a | 8.23  a | 125.28  a | 14.44  a | 40.04  b |
| MLE + FP | 4.12  b | 86.47  a | 8.69  a | 140.89  a | 14.86  a | 39.36  b |
| SE + MLE + FP | 4.99  a | 87.06  a | 8.55  a | 147.99  a | 15.58  a | 46.13  a |

Ctrl: control; RDCF: recommended dose of chemical fertilizer; SE: seaweed extracts; FP: microbial- based biostimulant of Fulzume-plus; MLE: moringa leaf extract.

Different letters in the same column indicate significant differences between means according to Duncan's new multiple range test at P≤ 0.05

The functions of the root and the shoot are closely integrated and each of them depending on the other. The root system takes up water and nutrients from the soil. The uptake and assimilation of materials from the soil requires energy and carbon skeletons which are largely provided from the shoot in the form of sucrose through the phloem. Thus, the root system relies on shoot-produced sugars. It may also be dependent on shoot-produced hormones to stimulate and regulate its growth. On the other hand, the shoot depends on the root system for supplying water and nutrients and for root-produced hormones such as gibberellins and cytokinins, which in turn stimulate shoot development. So, the functions of the root and the shoot are closely integrated and each is dependent on the other for its survival (Chesworth *et al.*, 1998).

Shoot growth characteristics of the plants such as plant shoot height, number of nodes per plant, stem diameter, plant leaf area, biomass accumulation and leaf chlorophyll intensity are important traits for determining plant yield in terms of quantity and quality. All these characters are strongly influenced by conventional and organic farming practices (Sulaiman, 2020). In our study, although the application of recommended dose of chemical fertilizer (RDCF) recorded the highest values of all studied characteristics related to the cucumber vegetative growth, but it was not differed significantly with the triple interaction between the biostimulants (SE + MLE + FP) which was used as an eco-friendly and alternative to the chemical fertilizer application. In addition, the majority treatments of individual and dual application between the biostimulans had a substantial impact on all studied traits, if compared to untreated plants (Table 4.2).

The improving vegetative growth characteristics due to the application of SE may be attributed to the auxins content in the SE which have an effective role in cell division and enlargement; this leads to increase the shoot height, nodes number, stem diameter, leaves area and finally chlorophyll intensity (Sarhan, 2011). SE also contains cytokinins in which induce the physiological activities, for example activating some enzymes that involved in photosynthesis, and increase the total chlorophyll in the leaves, this will positively reflect on the activity of photosynthesis process and consequently the synthesized materials positively reflected on the vegetative growth characteristics and roots growth as well (Shukla *et al.,* 2019; Chesworth *et al.*, 1998). Furthermore, this increase in shoot characteristics might also due to the content of macro and miconutrients in SE, which are very essential for plant growth and development. In addition, SE stimulates nutrients translocation and metabolism in the plants (Rasouli *et al.,* 2022). For these reasons, soil application of SE in our study caused significant increase of cucumber shoot height by 26.0%, number of nodes per plant by 36.9%, stem diameter by 28.1%, plant leaf area by 51.9% and leaf chlorophyll intensity by 20.2%, if compared to untreated plants.

Furthermore, moringa leaves are a potential source of natural antioxidants, growth enhancing compounds, amino acids, macro and micronutrients in its leaves extracts (MLE) (Yasmeen *et al.,* 2014). All these characteristics of MLE improved cucumber plant growth characteristics. The other reasons for the increase of vegetative growth traits of the cucumber plants by foliar application of MLE compared to control are probably due to improve the plants root growth (Table 4.1) and enhancing the nutrients uptake particularly N, P, K and Zn (Table 4.3) which ultimately positively improved total chlorophyll intensity (22.8%) and the other studied vegetative growth characteristics such as plant shoot height (36.5%), number of nodes per plant (22.0%), stem diameter (23.9%) plant leaf area (34.6%) and ultimately increased leaf dry matter percentage (44.3%).

Also, FP treatment caused significant increase in chlorophyll intensity in the cucumber leaves by 17%, this may be due to the impacts of the beneficial bacteria (*Bacillus spp* and *Pseudomonas spp*) that present in this product on increasing the ACC-deaminase enzyme in the treated plants which it slows down chlorophyll degradation (Mohamed and Gomaa, 2012); or probably due to the improvement of N and Fe concentrations in the leaves as a result of FP application (Table 4.3); in which these nutrients play a vital role in production of photosynthetic pigments and consequently the chlorophyll intensity increased (Taiz and Zeiger, 2002). On the other hand, FP play significant role in increasing increased Zn concentration in the cucumber leaves as shown in (Table 4.3), which is an essential nutrient that is directly involved in the biosynthesis of indole-3-acetic acid (IAA) in the plants (Kumar *et al.,* 2016). IAA is as important signal molecule in the regulation of plant growth and development, influencing many cellular plant processes, such as shoot and root cell division and elongation. Additionally, *Bacillus* and *Pseudomonas* can produce ammonia ($NH_4$) which is reported as another important trait of PGPR that may indirectly influence the plant growth (Joseph *et al.,* 2007). In addition, the secretion of ACC deaminase by PGPR inhibits ethylene biosynthesis in crop plants and promotes plant growth (Pourbabaee *et al.,* 2016). For these reasons, in our study cucumber plants shoot height increased by 25.7%, nodes number by 27.8%, plant leaf area by (37.2%) and ultimately dry matter percentage in the leaves by 47.1% as a result of FP treatment. Furthermore, among the biostimulants treatments, the triple interaction between the biostimulants (SE + MLE + FP) exhibited the maximum values of the all-vegetative growth measurements, after the treatment of (RDCF) (Table 4.2). This may be due to the effects of each (SE), (MLE) and (FP) that used in this combination on improving the cucumber plants root system characteristics (Table 4.1) as well as enhancing the nutrient status of the plants (Table 4.3).

## 4.3 Effect of Biostimulants on Concentration of Some Macro and Micro Nutrients in Cucumber Leaves

### 4.3.1 Nitrogen (N) (mg/gm)

The results in Table (4.3) show that using non- microbial and bacterial- based biostimulants played a significant role in improving N concentration in cucumber leaves. Furthermore, cucumber plants that fertilized with the recommended dose of chemical fertilizers (RDCF) gave the highest concentration of N (28.7%); but it was not differ significantly with all biostimulant treatments except the application of (SE). Moreover, all of the biostimulant treatments were significantly superior to control, which had the lowest concentration of N in the leaves (11.7%). In comparison to control, the biostimulant treatments of (SE), (MLE) and (FP) markedly raised N content in the cucumber leaves by 85.5, 133.3, and 106.0%, respectively. Also, increasing percentage for N content due to combinations of these biostimulants (SE + MLE), (SE + FP), (MLE + FP) and (SE + MLE + FP) were 133.3, 131.6, 123.1 and 130.8%, respectively.

### 4.3.2 Phosphorus (P) (mg/gm)

The results of Table (4.3) indicate that the addition of biostimulants significantly affected P concentration in the cucumber leaves, in which all of the treatments were significantly superior to the control, although they differed statistically among them. The highest P concentration (7.8%) was observed by the application of recommended dose of chemical fertilizer (RDCF), which it was not differ significantly with the treatments of (SE + MLE) and (SE + MLE + FP) which were registered 7.1 and 6.8% respectively. In addition, the lowest concentration of P was observed by the control (1.9%). In comparison to control, the biostimulant treatments of (SE), (MLE) and (FP) significantly improved the concentration of P in the leaves by 168.4, 173.7, and 168.4%, respectively. Also, increasing percentage for the combinations between the bistimulants (SE + MLE), (SE + FP), (MLE + FP) and (SE + MLE + FP) were reached 273.7, 221.1, 147.4 and 257.9%, respectively.

### 4.3.3 Potassium (K) (mg/gm)

Data in Table (4.3) illustrate that K concentration in the cucumber leaves was affected significantly by using non- microbial and bacterial- based biostimulants. The highest concentration was resulted from the application of (FP) (26.5%), but all the other biostimulant treatments, except the application of (SE), did not differ significantly with it. In addition, the lowest concentration of the K was recorded by control (11.3%). In comparison to control, individual applying of (SE), (MLE) and (FP) caused significant increase of K concentration by 90.3, 121.2 and 134.5%, respectively. Also, the combination effects of these biostimulants in the treatments of (SE + MLE), (SE + FP), (MLE + FP) and (SE + MLE + FP) significantly increased this trait by 91.2, 127.4, 123.0 and 125.7%, respectively.

*Chapter Four*                                                                                              *Results and Discussion*## 4.3.4 Iron (Fe) (mg. kg-1)

The results in Table (4.3) show that providing non- microbial and bacterial- based biostimulants played a significant role in improving Fe concentration in cucumber leaves. However, cucumber plants that fertilized with the recommended dose of chemical fertilizers (RDCF) gave the highest concentration of Fe (47.00 mg kg$^{-1}$), but it was not differ significantly with all biostimulant treatments except the applications of (MLE) and (FP). Moreover, control had the lowest content of Fe in the leaves (22.98 mg kg$^{-1}$). Among the individual application of the biostimulants, only applying (SE) significantly led to increase the content of Fe in the leaves and the percentage of the increase was 71.6%. In addition, increasing percentage for Fe concentration as a result of combinations of the biostimulants (SE + MLE), (SE + FP), (MLE + FP) and (SE + MLE + FP) were 81.5, 86.6, 62.4 and 90.7%, respectively.

## 4.3.5 Zinc (Zn) (mg. kg$^{-1}$)

The Zn concentration in the cucumber leaves was dramatically increased by all biosimulant treatments (Table 4.3). The highest Zn content is found in cucumber plants that have been treated with the bacterial-based biostimulant (FP) (71.00 mg kg$^{-1}$). In addition, it was not different significantly with the treatments of (SE), (SE + MLE), (MLE + FP) and (SE + MLE + FP). On the other hand, the control plants showed minimum content of Zn (36.73 mg kg$^{-1}$). In comparison to control, the biostimulant treatments of (SE), (MLE) and (FP) significantly improved Zn content in the cucumber leaves by 57.0, 46.8, and 93.3%, respectively. Also, increasing percentage for Zn content due to combinations of these biostimulants (SE + MLE), (SE + FP), (MLE + FP) and (SE + MLE + FP) were 72.9, 33.1, 60.8 and 77.1%, respectively.

**Table 4. 3: Effect of biostimulants on concentration of some macro and micro nutrients in cucumber leaves**

| Treatments | N (gm/kg$^{-1}$) | P (gm/kg$^{-1}$) | K (gm/kg$^{-1}$) | Fe (mg kg$^{-1}$) | Zn (mg kg$^{-1}$) |
|---|---|---|---|---|---|
| Ctrl. | 11.7 c | 1.9 e | 11.3 c | 22.98 b | 36.73 d |
| RDCF | 28.7 a | 7.8 a | 26.2 ab | 47.00 a | 52.41 bc |
| SE | 21.7 b | 5.1 cd | 21.5 b | 39.44 a | 57.65 abc |
| MLE | 27.3 a | 5.2 cd | 25.0 ab | 26.98 b | 53.91 bc |
| FP | 24.1 ab | 5.1 cd | 26.5 a | 26.90 b | 71.00 a |
| SE + MLE | 27.3 a | 7.1 ab | 21.6 ab | 41.72 a | 63.50 abc |
| SE + FP | 27.1 a | 6.1 bc | 25.7 ab | 42.88 a | 48.90 cd |
| MLE + FP | 26.1 ab | 4.7 d | 25.2 ab | 37.32 a | 59.06 abc |
| SE + MLE + FP | 27.0 a | 6.8 ab | 25.5 ab | 43.83 a | 65.06 ab |

Ctrl: control; RDCF: recommended dose of chemical fertilizer; SE: seaweed extracts; FP: microbial- based biostimulant of Fulzume-plus; MLE: moringa leaf extract.

Different letters in the same column indicate significant differences between means according to Duncan's new multiple range test at P≤ 0.05

Concentration of macro and micronutrients in plant leaves plays crucial roles in the plant growth and development. This is due to their participations in physiological activities and biosynthesis of proteins, enzymes, nucleic acids, pigments, hormones and antioxidants (Taiz and Zeiger, 2002). Therefore, the concentration of essential nutrients in plant tissues affects vegetative and reproductive growth as well as plant yield and yield components. In addition, nutrients uptake and their concentrations in plant tissues are strongly correlated to the conventional and organic farming management practices (Sulaiman and Sadiq, 2020a).

In the current study, even though the application of recommended dose of chemical fertilizer (RDCF) recorded the highest concentration of most the measured macro and micronutrients in the cucumber leaves, but the application of non-microbial and microbial biostimulants, also have a significant impact on improving nutrient status in the leaves in which some of the biostimulant treatments did not differ significantly with (RDCF) (Table 4.3). The reasons for the positive effect of SE on increasing the concentration of nutrients may be due to the fact that SE modified the root system characteristics which increased significantly number of lateral roots per plant, average lateral root length and diameter (Table 4.1). It means absorption surface area improved by SE (Battacharyya *et al.,* 2015). Furthermore, SE contains a component called kahydrin, a derivative of vitamin K1, which modifies the plasma membrane proton pumps and enhances the secretion of $H^+$ ions into the apoplast, leading to rhizosphere acidification. This condition alters the soil redox state and metal ion solubility, increasing their availability to the plant (Lüthje and Böttger, 1995). In addition to the fact that SE itself contains amount of macro and micronutrients that directly affected on improving these nutrients in the cucumber leaves by enhancing nutrients translocation and metabolism in the plants (Rasouli *et al.,* 2022). For these mentioned reasons SE increased N, P, K, Fe and Zn concentrations by 85.5, 168.4, 90.3, 71.6 and 57.0%, respectively if compared to untreated plants.

Moreover, enhancing nutrients concentration in the cucumber leaves due to the foliar application of MLE may be attributed to the macro and micronutrients that present in this extract, such as N, P, K, Ca, B, Mg, Cu, Zn, Mn, Na and Fe (Yasmeen *et al.,* 2014). The presence of mineral nutrients in MLE could be an option for supplementing the nutritional requirements of horticultural crops (Zulfiqar *et al.,* 2020). In addition, increased nutrients in cucumber leaves may be due to the plant growth substances that contained in the MLE, especially IAA, which is considered a very important stimulator of root growth (Taiz and Zeiger, 2002) in which significantly increased the number of lateral roots per plant, lateral roots length and diameter (Table 4.1). This means increasing the absorption area of the roots, which led to an improvement in the roots ability to absorb nutrients from the soil, and then reflected on increasing the concentration of nutrients in the leaves. So, in this study the concentration of N, P, K and Zn increased substantially as a result of MLE application by 133.3, 173.7, 121.2 and 46.8% respectively, in comparison to untreated plants. However, the increase in Fe concentration did not reach the significant level. Furthermore, beneficial bacteria such as *Bacillus* and *Pseudomonas* species can convert the complex form of essential nutrients in the soil to a simple available form which easily accessible by the plant roots (Kang *et al.,* 2015; Kuan *et al.,* 2016). The available form of N in soil is limited, which slows plant growth in natural habitats. Some of the PGPR release ammonia from nitrogenous organic matter (Hayat *et al.,* 2010). In this regard, Radhakrishnan *et al.,* (2017) stated that some of the *Bacillus* spp. have the nifH gene and produce nitrogenase, which can fix atmospheric $N_2$ and provide it to plants in order to enhance plant growth and productivity. Additionally, the secretion of phosphatases and low molecular weight organic acids, such as gluconic acid, from *Bacillus* spp. acidifies the surrounding environment to facilitate the conversion of inorganic phosphate into free mineral from phosphate (Kang *et al.,* 2015).

Furthermore, some species of *Bacillus* and *Pseudomonas* are capable of solubilizing K and Zn through the secretion and production of organic acids (Hafeez and Hassan, 2012; Pramanik *et al.,* 2019). In addition, the soil of our experiment is characterized by its relatively high pH value and high content of calcium carbonate ($CaCO_3$) (Table 3.1), which they have a significant impact on decreasing nutrients availability in the soil solution. Therefore, in this study the application of FP as a bacterial-based biostimulant that contains (*Bacillus subtilis* and *Pseudomonas putida*) resulted in a substantial increase of N, P, K and Zn concentration in cucumber leaves by 106.0, 168.4, 134.5 and 93.3%, respectively. However, the increase in Fe concentration did not reach the significant level.

## 4.4 Effect of Biostimulants on Reproductive Growth and Yield Component Characteristics of Cucumber

### 4.4.1 Number of flowers per plant

According to the findings in (Fig. 4.1), neither the application of chemical fertilizer (RDCF) nor the biostimulants of (SE, MLE and FP), individually or in combinations, significantly affected the number of cucumber flowers per plant when compared to the control.

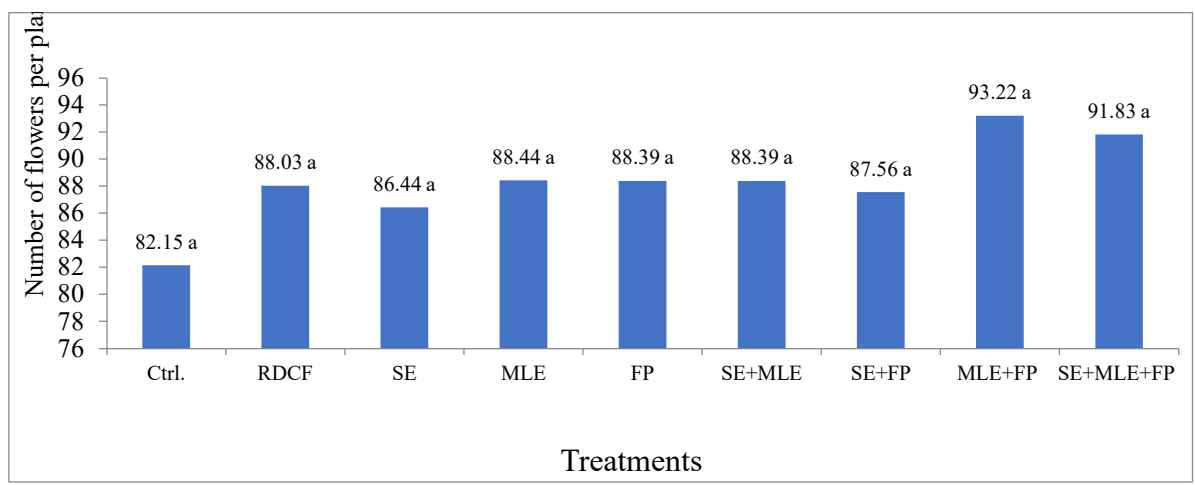

**Figure 4. 1: Effect of biostimulants on number of flowers per plant.**

Ctrl: control; RDCF: recommended dose of chemical fertilizer; SE: seaweed extracts; FP: microbial- based biostimulant of Fulzume-plus; MLE: moringa leaf extract.

Different letters indicate significant differences between means according to Duncan's new multiple range test at P≤ 0.05

### 4.4.2 Number of aborted flowers per plant

The average number of aborted flowers per plant was affected significantly by applying most of the biostimulant treatments (Fig. 4.2). The lowest number was resulted from the RDCF (26.75 flower plant$^{-1}$), but it was not differed significantly with all the biostimulant treatments, except the application of FP. On the other hand, the highest number was recorded by the control (43.33 flower plant-1). Applying SE and MLE resulted in a significant reduction of the abortion phenomenon by 20.8 and 25.9%, respectively when compared to control. Additionally, this phenomenon was significantly decreased by 20.8, 21.3, 23.4 and 28.5 % as a result of the application of (SE + MLE), (SE + FP), (MLE + FP) and (SE + MLE + FP), respectively.

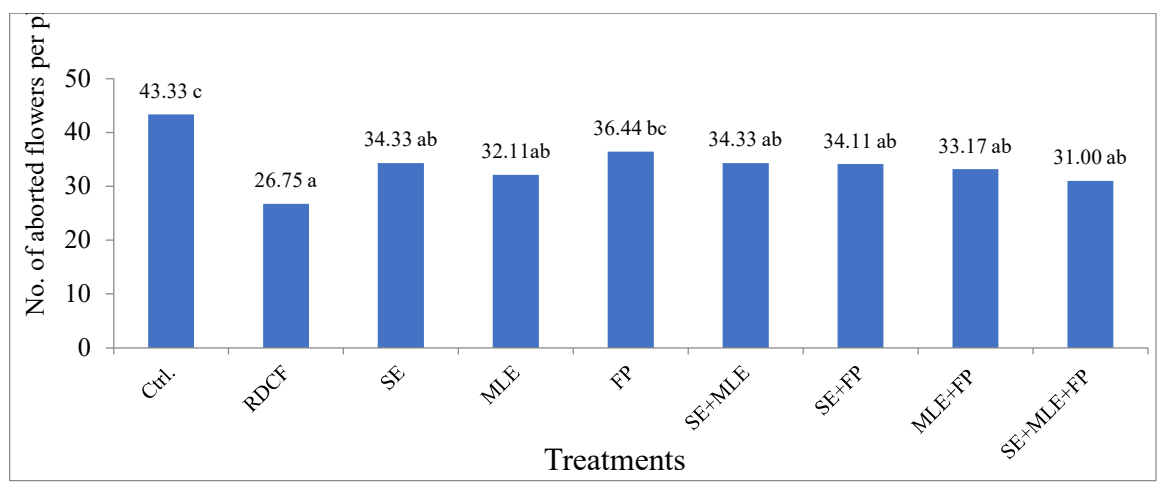

**Figure 4. 2 Effect of biostimulants on number of aborted flowers per plant.**

Ctrl: control; RDCF: recommended dose of chemical fertilizer; SE: seaweed extracts; FP: microbial- based biostimulant of Fulzume-plus; MLE: moringa leaf extract.

Different letters indicate significant differences between means according to Duncan's new multiple range test at P≤ 0.05

### 4.4.3 Fruit set percentage (%)

The influence of non-microbial biostimulants SE and MLE as well as bacterial- based biostimulants of Fulzym- plus FP on cucumber fruit set percentage is shown in (Fig. 4.3). Even though the highest percentage of the cucumber fruit set (69.66%) was observed by the application of chemical fertilizers RDCF, but it was not statistically superior to the biostimulant treatments of (MLE + FP) and (SE + MLE + FP) which were recorded 64.58 and 66.33% respectively. In addition, cucumber plants grown without adding any kinds of biostimulants (Control) had the lowest percentage (47.18%). In comparison to control, the biostimulant treatments of SE, MLE and FP significantly enhanced cucumber fruit set by 27.8, 35.0, and 24.9%, respectively. Also, increasing the fruit set was due to combinations of these biostimulants (SE + MLE), (SE + FP), (MLE + FP) and (SE + MLE + FP) were 29.1, 30.9, 36.9 and 40.6%, respectively.

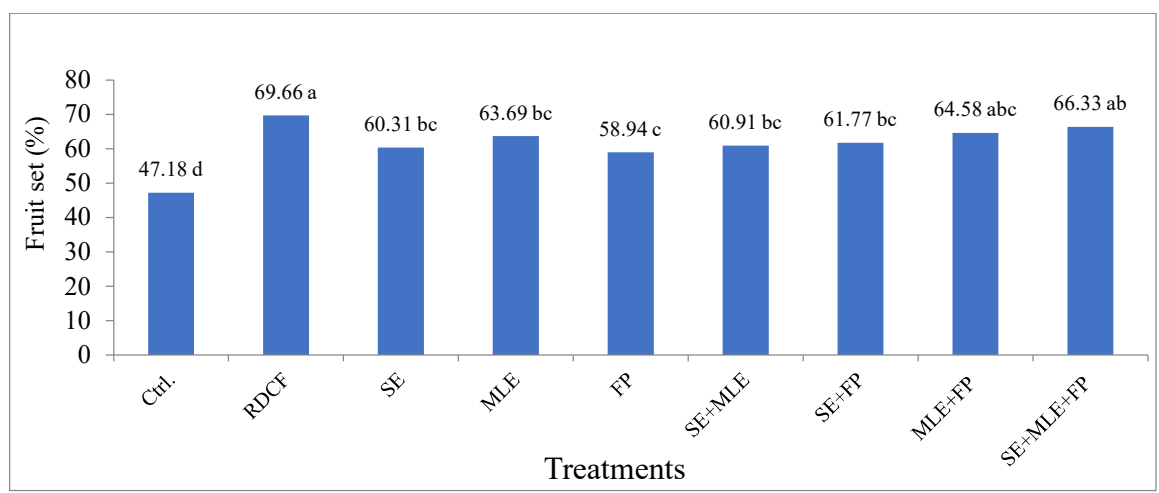

**Figure 4. 3: Effect of biostimulants on cucumber fruit set (%)**

Ctrl: control; RDCF: recommended dose of chemical fertilizer; SE: seaweed extracts; FP: microbial- based biostimulant of Fulzume-plus; MLE: moringa leaf extract.

Different letters indicate significant differences between means according to Duncan's new multiple range test at P≤ 0.05

### 4.4.4 Number of fruits per plant

All the biostimulant treatments were substantially affected the number of cucumber fruits per plant (Fig. 4.4). Despite the highest fruit number was produced by the cucumber plants that fertilized with recommended dose of chemical fertilizer (RDCF) (61.28 fruit plant$^{-1}$), it was not varied significantly with majority of the biostimulant treatments including: (MLE), (SE + MLE), (SE + FP), (MLE + FP) and (SE + MLE + FP). Furthermore, the control plants recorded the lowest fruits number per plant with an average of (38.82 fruit plant$^{-1}$). Individual applications of (SE), (MLE), and (FP) significantly increased cucumber fruits number by 34.2, 45.1, and 33.8%, respectively, as compared to control. Additionally, the combined effects of these biostimulants considerably improved this trait by 39.3, 37.7, 54.7, and 56.7%, respectively, for the treatments of (SE + MLE), (SE + FP), (MLE + FP), and (SE + MLE + FP).

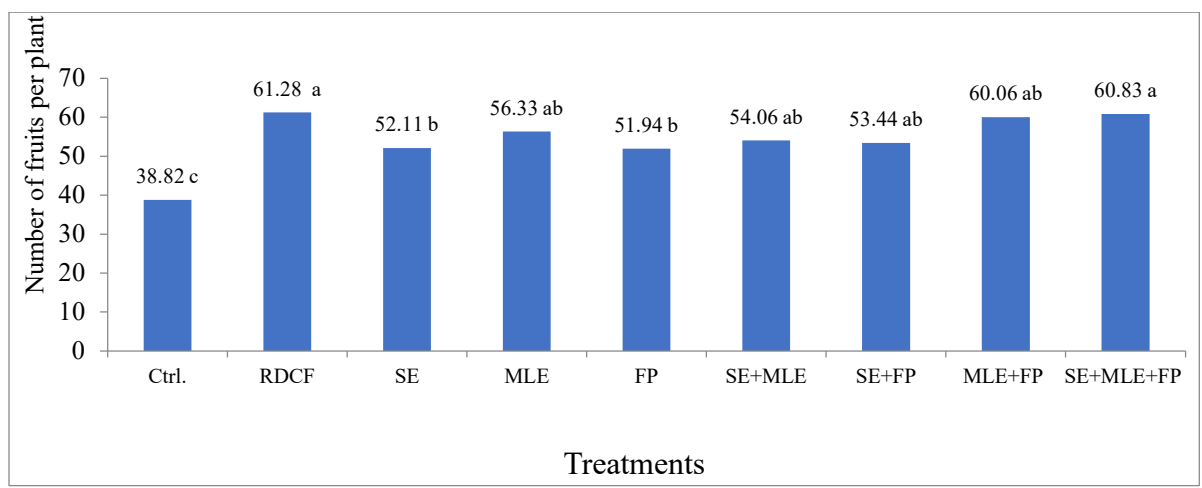

**Figure 4. 4 Effect of biostimulants on number of cucumber fruits per plant**

Ctrl: control; RDCF: recommended dose of chemical fertilizer; SE: seaweed extracts; FP: microbial- based biostimulant of Fulzume-plus; MLE: moringa leaf extract.
Different letters indicate significant differences between means according to Duncan's new multiple range test at P≤ 0.05

## 4.4.5 Fruit weight (g)

The influence of non-microbial biostimulant treatments SE and MLE as well as bacterial- based biostimulants of Fulzym- plus (FP) on average fruit weight is shown in (Fig. 4.5). The highest fruit weights (97.30, 97.09, 96.79, 96.14 and 95.13 g) were obtained by the applications of SE, RDCF, FP, control and (SE + MLE + FP), respectively. However, the application of (SE + MLE) recorded the lowest average fruits weight with an average of (86.30 g) although it was not significantly different from (MLE) and (SE + FP).

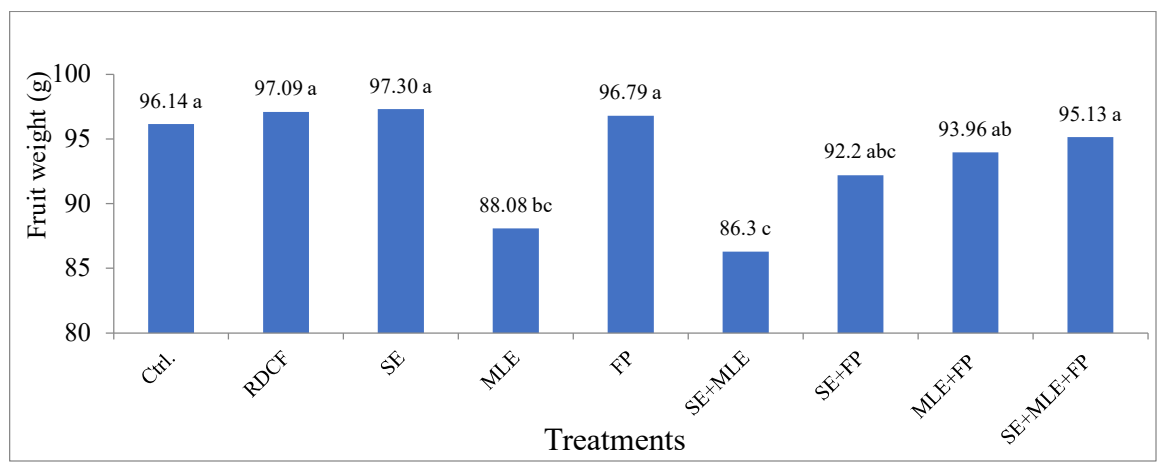

**Figure 4. 5: Effect of biostimulants on average fruit weight (g).**

Ctrl: control; RDCF: recommended dose of chemical fertilizer; SE: seaweed extracts; FP: microbial- based biostimulant of Fulzume-plus; MLE: moringa leaf extract.

Different letters indicate significant differences between means according to Duncan's new multiple range test at P≤ 0.05

### 4.4.6 Early Yield (kg plant$^{-1}$)

The results in (Fig. 4.6) show the impact of the biostimulant treatments on the early yield of cucumber plants. Despite the highest early yield was produced by the cucumber plants that fertilized with recommended dose of chemical fertilizer (RDCF) (0.96 kg plant$^{-1}$), it was not different significantly with most of the biostimulant treatments, including: SE, FP, (MLE + FP) and (SE + MLE + FP). Furthermore, control plants registered the lowest amount of early yield per plant with an average of (0.56 kg plant$^{-1}$). Among the bistimulant treatments, only the applications of (MLE + FP) and (SE + MLE + FP) were significantly superior to the control, and increasing percentage reached 58.9 and 60.7%, respectively.

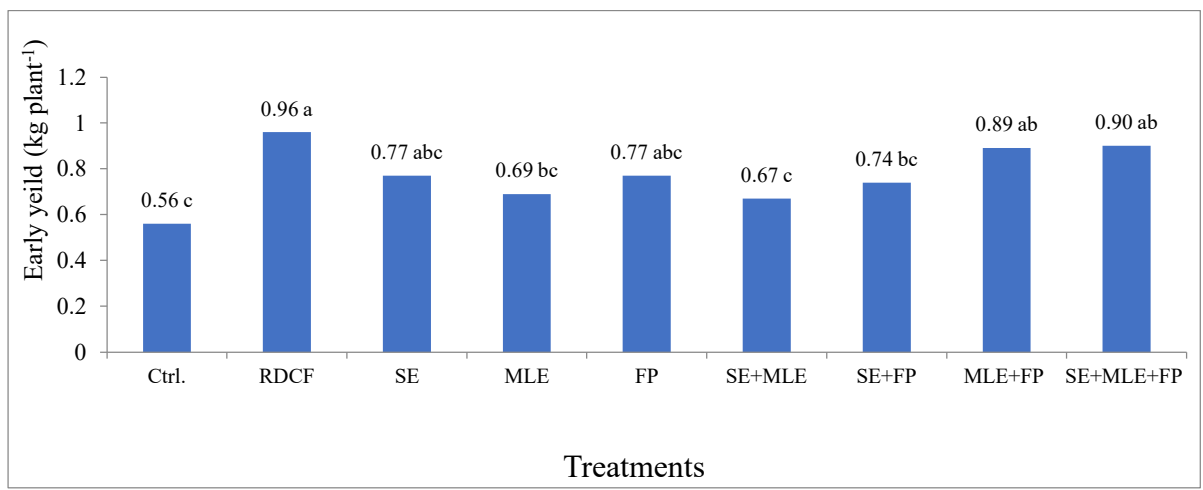

**Figure 4. 6 Effect of biostimulants on early yield (kg plant$^{-1}$).**

Ctrl: control; RDCF: recommended dose of chemical fertilizer; SE: seaweed extracts; FP: microbial- based biostimulant of Fulzume-plus; MLE: moringa leaf extract.

Different letters indicate significant differences between means according to Duncan's new multiple range test at P≤ 0.05

### 4.4.7 Plant yield (kg plant$^{-1}$)

All the biostimulant treatments were significantly impacted on the cucumber plants yield (Fig. 4.7). Although the maximum yield was produced by the cucumber plants that fertilized with recommended dose of chemical fertilizer (RDCF) (5.95 kg plant$^{-1}$), it was not different significantly with the biostimulant treatments of (MLE + FP) and (SE + MLE + FP). Which they recorded 5.64 and 5.79 kg. plant$^{-1}$ respectively. Furthermore, the control plants recorded the lowest yield per plant (3.70 kg plant$^{-1}$). Individual applications of SE, MLE, and FP markedly increased cucumber productivity by 37.0, 34.1, and 35.9%, respectively, as compared to control. Additionally, the combined effects of these biostimulants considerably enhanced this trait by 26.8, 33.2, 52.4, and 56.5%, respectively for the treatments of (SE + MLE), (SE + FP), (MLE + FP), and (SE + MLE + FP).

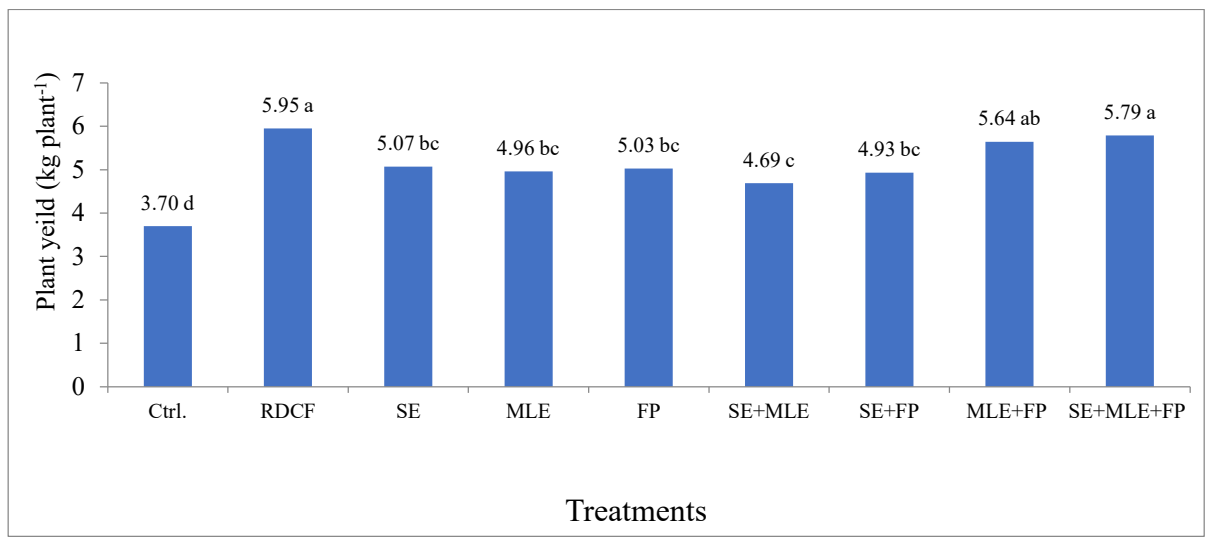

**Figure 4. 7 Effect of biostimulants on cucumber plant yield (kg plant$^{-1}$)**

Ctrl: control; RDCF: recommended dose of chemical fertilizer; SE: seaweed extracts; FP: microbial- based biostimulant of Fulzume-plus; MLE: moringa leaf extract.
Different letters indicate significant differences between means according to Duncan's new multiple range test at P≤ 0.05

The plants reproductive growth characteristics such as number of total flowers per plant, flower abortion phenomenon and fruit set percentage; also, plant yield component traits particularly number of fruits per plant and average fruit weight, as well as their balance, are the most significant features that have a direct relationship to the quantity of plant output (Mahmood *et al.,* 2021). These mentioned characteristics are most affected by agronomic services including nutrient management practices whether in conventional or organic farming (Sulaiman and Sadiq, 2020a).

The results of this study revealed that the application of non-microbial biostimulants (SE and MLE) and bacterial- based biostimulant of (FP), individually and their combinations had a significant impact on improving some reproductive growth and plant yield characteristics of cucumber plants if compared to untreated plants. Furthermore, using the three biostimulants together in the treatment of (SE + MLE + FP) did not different substantially with (RDCF) in all studied reproductive growth and yield component traits (Figs. 4.1 – 4.7).

The reasons for improving cucumber flowering development and plant yield by SE application may be linked with a number of physiological and biochemical mechanisms, including the elicitation of enzymes involved in carbon and nitrogen metabolic pathways, the Krebs cycle and glycolysis as well as the stimulation of phytohormones (Colla *et al.,* 2017 ;La Bella *et al.,* 2021). in addition to the enhancement of nutrients acquisition and accumulation of treated plants through root morphology alterations (Table 4.1 and 4.3). Furthermore, SEs are high in bioactive components such as macro and micronutrients, essential fatty acids, amino acids, vitamins, auxins and cytokinins, which boosted cellular metabolism in treated plants, resulting in an improvement of plant growth and productivity (Hamouda *et al.*, 2022). Thereby, in our study as a result of the SE application, the reproductive traits and productivity of the cucumber plants such as decreasing number of aborted flowers per plant fruit set percentage, number of fruits per plant, and plant yield were significantly improved by 20.8, 27.8, 34.2 and 37.0% respectively if they compared to the untreated plants (Figs. 4.2, 4.4 and 4.7).

The ability of MLE to improve some of the measured flowering growth as well as plant yield characteristics of cucumber plants is attributed to the bioactive compounds present in this extract. These compounds or active ingredients can improve certain key physiological, biochemical and molecular processes. These mechanisms enhance the availability of nutrients in the plants and positively affect the growth and yield parameters, as well as the nutritional quality of the fruits (Mashamaite *et al.*, 2022). The application of MLE also enhances photosynthesis and further improves the metabolism of carbon and nitrogen (Yasmeen *et al.*, 2013). Furthermore, the presence of phytohormones in the MLEs plays a crucial role in cell division, resulting in cell multiplication and general cell enlargement or elongation, ultimately leading to the improved growth and yield of crops (Elzaawely *et al.*, 2017). For mentioned above reasons, in this study foliar application of MLE significantly increased fruit set percentage by 35.0% as a result of decreasing the number of aborted flowers per plant by 25.9% (Figures 4.3 and 4.2). So, number of fruits per plant improved by 45.1%, and finally the plant yield increased by 34.1% compared to untreated plants (Figs. 4.4 and 4.7).

Furthermore, using beneficial microorganisms in horticulture as plant biostimulant represents an eco-friendly alternative to the chemical products, which is also a new approach to the organic farming practices (Inculet *et al.*, 2019; Stefan *et al.*, 2013). In our study, we used FP which is a commercial bacterial-based biostimulant that contains beneficial bacteria of *Bacillus subtilis* and *Pseudomonas putida*. These bacteria are the most predominant group in the plant growth-promoting rhizobacteria (PGPR). Cucumber plants that treated with this product improved significantly plants root growth (Table 4.1), chlorophyll intensity and some of the other vegetative growth characteristics (Table 4.2), and nutrients concentrations in the leaves such as N, P, K and Zn (Table 4.3). All mentioned changes were led to significantly improve in the fruit set percentage by 24.9% (Figure 4.3), thereby the number of fruits per plant and plant yield were increased significantly as a result of FP application by 33.8 and 35.9% respectively, if compared to untreated plants (Figs. 4.4 and 4.7). The synergism effects of the three mentioned biostimulants in the treatment of (SE +MLE + FP) could be the reason for its superiority to the control and some of the other biostimulant treatments.

## 4.5 Effect of Biostimulants on Cucumber Fruit Quality Characteristics

### 4.5.1 Fruit length (cm)

The influence of the non-microbial biostimulants as well as the bacterial- based biostimulant treatments on average fruit length is shown in (Fig. 4.8). The longest fruit (17.17 cm) was recorded by the application of (SE + MLE + FP). However, it was not different substantially from (RDCF), (SE) and (MLE + FP). The biostimulant treatments of (MLE), (FP), (SE + MLE), and (SE + FP) were ranked moderate fruit length with an average of 15.94, 15.44, 15.56 and 15.86 cm, respectively. Additionally, control plants produced the shortest fruits (14.19 cm). As compared to control, increasing percentage of the fruit length as a result of the biostimulant applications were 13.5% for (SE), 12.3% for (MLE), 8.8% for (FP), 9.7% for (SE + MLE), 11.8% for (SE + FP), 15.1% for (MLE + FP) and 21.0 for (SE + MLE + FP).

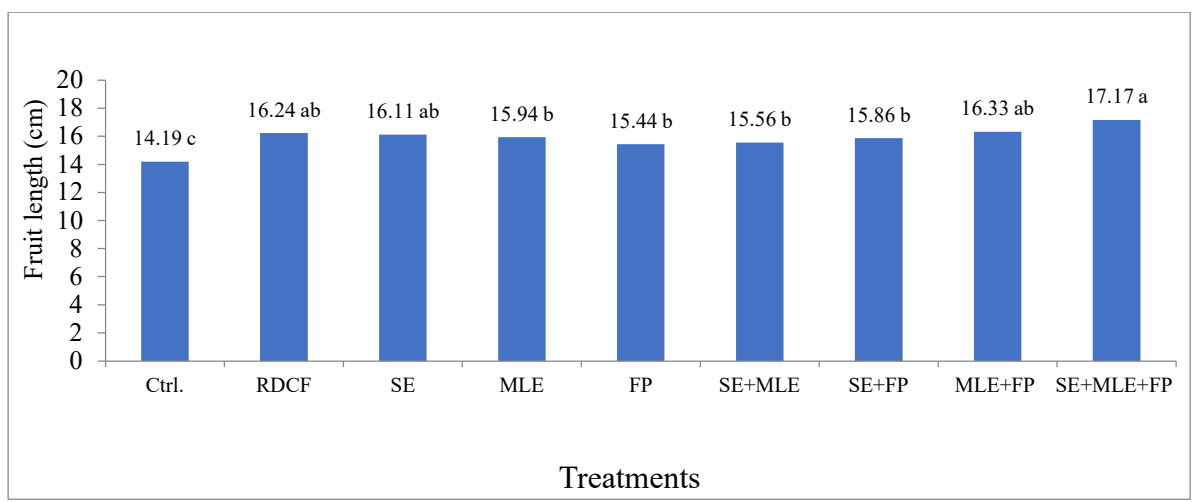

**Figure 4. 8: Effect of biostimulants on cucumber fruit length (cm)**

Ctrl: control; RDCF: recommended dose of chemical fertilizer; SE: seaweed extracts; FP: microbial- based biostimulant of Fulzume-plus; MLE: moringa leaf extract.
Different letters indicate significant differences between means according to Duncan's new multiple range test at P≤ 0.05

### 4.5.2 Fruit diameter (mm)

Results in (Fig. 4.9) show the effect of the biostimulants treatments on average fruit diameter. The maximum diameters (33.39, 32.68 and 32.63 mm) were obtained by the applications of (RDCF), (SE + FP) and (SE + MLE + FP), respectively. However, they did not differ significantly from the other biostimulant treatments. On the other hand, the smallest fruits diameter (28.33 mm) was produced by the control plants. The biostimulant treatments of (SE + FP) and (SE + MLE + FP), considerably increased this trait by 15.4 and 15.2%, respectively.

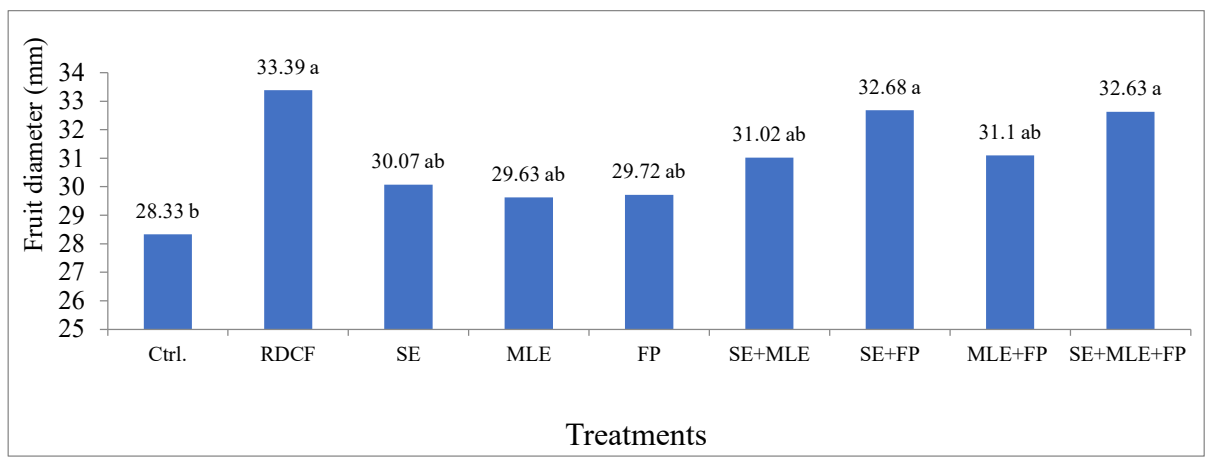

**Figure 4. 9:** Effect of biostimulants on cucumber fruit diameter (mm)

Ctrl: control; RDCF: recommended dose of chemical fertilizer; SE: seaweed extracts; FP: microbial- based biostimulant of Fulzume-plus; MLE: moringa leaf extract.

Different letters indicate significant differences between means according to Duncan's new multiple range test at P≤ 0.05

## 4.5.3 Fruit dry matter (%)

The impact of the non-microbial as well as the bacterial- based biostimulant treatments on cucumber fruits dry matter percentage is shown in (Figure 4.10). All the cucumber plants that treated with the biostimulants did not differ significantly with the plants that fertilized with the recommended dose of chemical fertilizer (RDCF) in this trait. In addition, all the biostimulant treatments, except the (SE + MLE), were markedly increased fruits dry matter percentage as compared to control. Furthermore, the increasing percentage as a result of the biostimulant applications reached 52.9% for (SE), 62.3% for (MLE), 50.9% for (FP), 62.6% for (SE + FP), 46.5% for (MLE + FP) and 64.9% for (SE + MLE + FP), as compared to control.

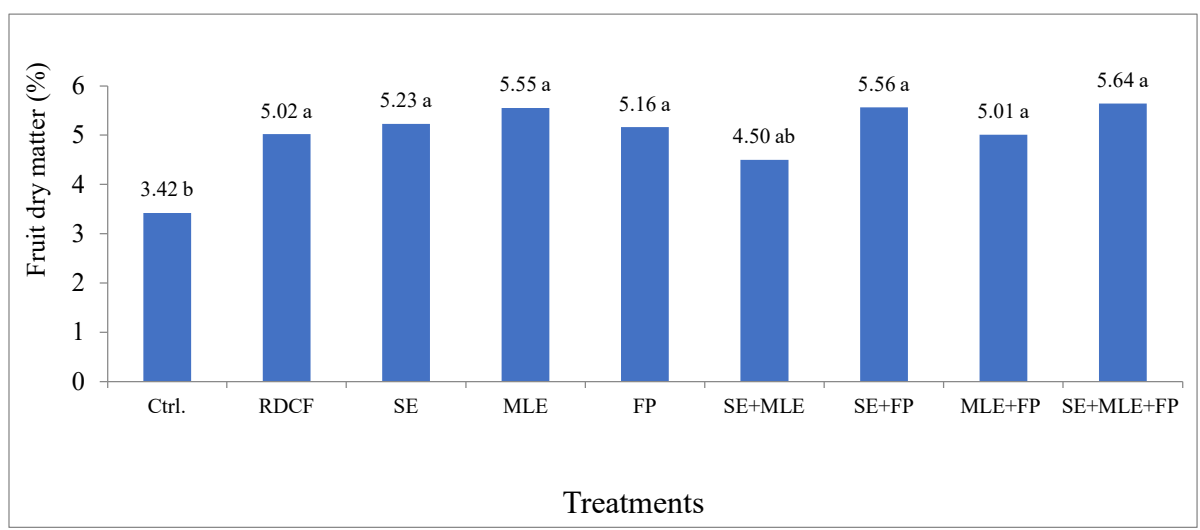

**Figure 4. 10:** Effect of biostimulants on cucumber fruit dry matter (%).

Ctrl: control; RDCF: recommended dose of chemical fertilizer; SE: seaweed extracts; FP: microbial- based biostimulant of Fulzume-plus; MLE: moringa leaf extract.

Different letters indicate significant differences between means according to Duncan's new multiple range test at P≤ 0.05

### 4.5.4 Fruit TSS (Brix)

The application of SE recorded the highest fruit TSS value (4.09˚ Brix) in which it was significantly superior only over the treatment of (SE + MLE) (3.51˚ Brix). The other treatments did not different significantly with the control (Fig. 4.11).

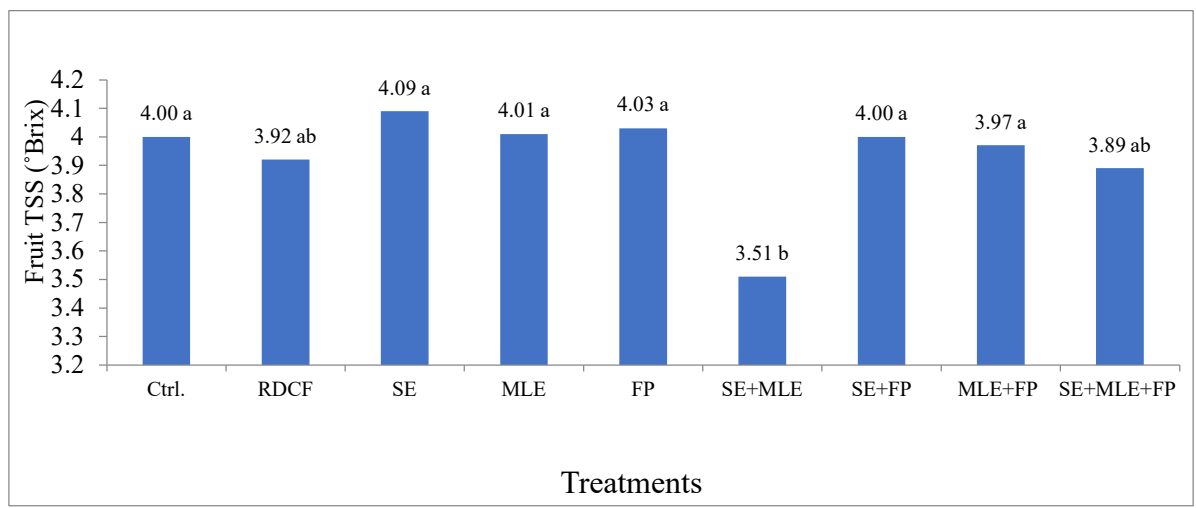

**Figure 4. 11: Effect of biostimulants on cucumber fruit TSS (˚Brix).**

Ctrl: control; RDCF: recommended dose of chemical fertilizer; SE: seaweed extracts; FP: microbial- based biostimulant of Fulzume-plus; MLE: moringa leaf extract.

Different letters indicate significant differences between means according to Duncan's new multiple range test at P≤ 0.05

The enhancement of vegetative growth by SE application (Table 4.2) reflected on the fruit growth, in which fruit length and dry matter percentage were significantly increased by 13.5 and 52.9% (Figs. 4.8 and 4.10). These might be resulted from more assimilates reaching the fruits or as a result of increasing fruit sink strength, which draws in more nutrients and assimilates (Abdel-Mawgoud *et al.,* 2010; Trejo Valencia *et al.,* 2018).

Furthermore, foliar spraying of MLE enhanced the length of cucumber fruits due to the presence of zeatin, the most common cytokinin in MLE (Nasir *et al.,* 2020). Zeatin accelerates cell division and cell enlargement during fruit development (Teribia *et al.,* 2016). Also, cytokinin improves sink capacity and photosynthate assimilation as a result of more expanded green areas in the leaf (Zwack and Rashotte, 2013). The higher fruit length and dry matter percentage in the fruits might also be associated with the role of minerals in MLE, especially K and Zn. K regulates the accumulation of sugars while Zn, a tryptophan precursor, is involved in IAA biosynthesis, which is required for the growth and development of fruits (Nasir *et al.,* 2016). Therefore, in our study cucumber fruits length and dry matter percentage were increased significantly by 12.3 and 62.3% due to foliar application of the MLE (Figs. 4.8 and 4.10).

Also, PGPR including *Bacillus* and *Pseudomonas* species have a vital role in improving cucumber fruit quality characteristics (Kafi *et al.*, 2021 and Zapata-Sifuentes *et al.*, 2022). This is may be through their influence on greater nutrients uptake from the soil by the inoculated plants; which these nutrients are considered an important factor in fruit growth (Kaloterakis *et al.*, 2021; Zapata-Sifuentes *et al.*, 2022). Therefore, in this study, the treatment of FP enhanced the fruit length and dry matter percentage of the cucumber fruits by 8.8 and 50.9%, respectively in comparison to untreated plants (Figs 4.8 and 4.10).

In addition, the combined effects of the three mentioned biostimulants in the treatment of (SE+MLE+FP) had a significant effect on cucumber fruit quality characteristics, in which it was not different substantially from the fruits that produced by plants that fertilized with recommended dose of chemical fertilizers (RDCF). This may be due to the individual effects of each biostimulants on improving root growth (Table 4.1) and vegetable growth (Table 4.2) as well as nutrient uptake (Table 4.3), which consequently had a positive effect on fruit growth. For these reasons, compared to untreated plants, this treatment increased fruit length by 21.0%, fruit diameter by 15.2% and finally fruit dry matter percentage by 64.9%.

### 4.6 Fruit Non-enzymatic Antioxidants:

#### 4.6.1 Fruit total phenolic contents (TPC) (µg GAE g$^{-1}$ FW)

The effect of the biostimulants on total phenolic content in the cucumber fruits is shown in (Table 4.4). Based on the outcomes, the application of the non-microbial biostimulant of (SE) recorded the highest content of TPC (66.51 µg GAE g$^{-1}$ FW), and it was significantly superior over the other treatments. In addition, the lowest TPC in the cucumber fruits (33.24 µg GAE g$^{-1}$ FW) was recorded by the plants that fertilized with recommended dose of chemical fertilizers (RDCF).

#### 4.6.2 Fruit total flavonoid content (TFC) (µg QE g$^{-1}$ FW)

Results in (Table 4.4) show the effect of the biostimulants treatments on total flavonoid contents in the cucumber fruits. The highest content (9.17 µg QE g$^{-1}$ FW) was found by the applying of (SE + MLE + FP); which it was significantly superior over the treatments of Ctrl., RDCF, SE and (SE + FP). On the other hand, the lowest TFC (3.86 µg QE g$^{-1}$ FW) was recorded by the cucumber plants that fertilized with recommended dose of chemical fertilizers (RDCF).

**Table 4. 4: Effect of biostimulants on fruit non-enzymatic antioxidant characteristics**

| Treatments | TPC (µg GAE g$^{-1}$ FW) | TFC (µg QE g$^{-1}$ FW) |
|---|---|---|
| Ctrl. | 40.14 b | 5.89 bc |
| RDCF | 33.24 b | 3.86 c |
| SE | 66.51 a | 5.80 bc |
| MLE | 36.94 b | 6.45 abc |
| FP | 44.13 b | 7.99 ab |
| SE + MLE | 40.05 b | 7.76 ab |
| SE + FP | 45.30 b | 5.94 bc |
| MLE + FP | 36.11 b | 7.81 ab |
| SE + MLE + FP | 38.10 b | 9.17 a |

Ctrl: control; RDCF: recommended dose of chemical fertilizer; SE: seaweed extracts; FP: microbial- based biostimulant of Fulzume-plus; MLE: moringa leaf extract.

Different letters in the same column indicate significant differences between means according to Duncan's new multiple range test at P≤ 0.05

Vegetable crops are important source of bioactive and antioxidant compounds, including phenolic acids and flavonoids, which are able to reduce the risk of cardiovascular diseases and specific forms of cancer (Rouphael *et al.,* 2018). The content of antioxidant compounds in different crops depended on several factors among which the most important is the farming system (organic or conventional) (Maggio *et al.,* 2013). In sustainable and organic farming, the use of different plant biostimulants, as eco-friendly products, is one of the most important strategies utilized to achieve high-quality fruits in terms of antioxidants content (Yakhin *et al.,* 2017). The plant biostimulants may activate the signaling pathway, which modulates the expression of gene-encoding enzymes associated with secondary metabolism (Machado *et al.,* 2014).

In our results, the significant increase of TPC in the cucumber fruits as a result of non-microbial biostimulant treatment of SE could be attributed to the stimulation of growth hormone biosynthesis and nutrient uptake by the roots; or possibly due to the improvement of polyphenol oxidase activity which increases the accumulation of phenolic compounds in the plant and hence in the fruits (Rasouli *et al.,* 2022). Therefore, compared to untreated plants, this treatment increased TPC in the fruits by 65.7%, while compared to RDCF increased by 100.1%. Concerning the individual effect of MLE and bacterial-based biostimulants on TPC, though recent studies have mentioned their roles in improving TPC in the fruits such as (Yuniati *et al.,* 2022; Ganugi *et al.,* 2021); while, their impacts were not statistically significant in our results (Table 4.4).

Regarding the TFC in the fruits, numerous studies have shown that SEs boost flavonoid contents in various vegetable crops, for instance (Lola-Luz *et al.,* 2013; Lola-Luz *et al.,* 2014), but the precise component in the SEs that triggers the phenylpropanoid and flavonoid pathways in plants is unknown yet (Fan *et al.,* 2011). Further study is necessary to clarify the mechanisms by which SEs cause significant increases in phytochemical content in plants and fruits (Lola-Luz *et al.,* 2013). Furthermore, MLE is an important source of natural antioxidants, such as flavonoid, quercetin and kaempferol, which positively improves antioxidant compounds in the fruits (Yuniati *et al.,* 2022). In addition, MLE contains minerals and vitamins that affect plant metabolic processes and consequently leads to increase TFC (Mehmood *et al.,* 2021).

# Conclusions and Recommendations

**Conclusions**

According to our results, the following points could be concluded:

1) Use of non-microbial biostimulants of (SE), (MLE) and bacterial-based biostimulants of (FP), individually and their combinations, had a significant impact on improving majority of the studied characteristics related to the cucumber plant growth, yield components and fruit quality.
2) The triple combination of the treatments (SE+MLE+FP) was more affective, which was not significantly different from the recommended dose of chemical fertilizers (RDCF) in most of studied characteristics.
3) The triple combination of the tested biostimulants maintained a better balance between quality and quantity of the cucumber products compared to the application of chemical fertilizers (RDCF).

**Recommendations**

Based on our conclusions, we recommend the following important points:

1) Using organic substances, different plant extracts and beneficial microorganisms as a powerful tool for organic agriculture to replace chemical fertilizers, which pollute the environment significantly.
2) Applying co-application of microbial and non-microbial biostimulants to improve plant growth, nutrients status, productivity and fruit quality traits of vegetable crops grown in organic farming system under greenhouse conditions.
3) Conducting studies on using organic substances, different plant extracts and other microbial biostimulants on the biosynthesis of biochemical compounds in the plants under greenhouse conditions to know their roles in the tolerance of crops to biotic and abiotic stresses.
4) Encourage farmers to produce vegetable crops organically in the greenhouses, and provide markets for their products by the government.